\documentstyle[11pt,aaspp4]{article}

\lefthead{K. Subramanian}
\righthead{Fluctuating Fields in Turbulent Dynamos}

\begin{document}

\title{Dynamics of fluctuating magnetic fields in turbulent dynamos 
incorporating ambipolar drifts}

\author{Kandaswamy Subramanian\altaffilmark{1}}
\affil{ Astronomy Centre, University of Sussex, Falmer, 
         Brighton BN1 9QH , UK.}
\affil{NCRA, TIFR,
        Poona University Campus, Ganeshkind, Pune 411007. India}


\altaffiltext{1}{On leave from 
       National Centre for Radio Astrophysics, TIFR,
        Poona University Campus, Ganeshkind, Pune 411007. India}


\begin{abstract}

Turbulence with a large magnetic Reyonolds number, generically
leads to rapidly growing magnetic noise over and above any mean field.
We revisit the dynamics of this fluctuating field, in homogeneous,
isotropic, helical turbulence. Assuming the turbulence to be Markovian,
we first rederive, in a fairly transparent manner,
the equation for the mean field, and 
corrected Fokker-Plank type equations for the magnetic
correlations. In these equations, we also 
incorporate the effects of ambipolar drift which would obtain
if the turbulent medium has a significant neutral component. 
We apply these equations to discuss a number of astrophysically 
interesting problems: 
(a) the small scale dynamo in galactic turbulence with a model Kolmogorov
spectrum, incorporating the effect of ambipolar drift; 
(b) current helicity
dynamics and the quasilinear corrections to the alpha effect;
(c) growth of the current helicity and large-scale magnetic fields 
due to nonlinear effects.

\end{abstract}



\section{Introduction}

The origin of large-scale cosmic magnetic fields remains at
present, a challenging problem. In a standard
paradigm, one invokes the dynamo action involving helical turbulence 
and rotational shear, to generate magnetic fields ordered
on scales much larger than the turbulence scale (cf. Moffat 1978, Parker 1979,
Zeldovich {\it et al.} 1983).
However, turbulent motions, with a
large enough magnetic Reynolds number (MRN henceforth),
can also excite a small-scale dynamo, which exponentiates fields correlated on
the tubulent eddy scale, at a rate much faster than the mean field
growth rate (Kazantsev 1968, Zeldovich {\it et al.} 1983 and see below).
A possible worry is that, these small-scale fields
can come to equipartition with the turbulence much before the 
large-scale field has grown appreciably and may then
interfere with the large-scale dynamo
action (cf. Kulsrud and Anderson, 1992 (KA)).
Indeed the efficiency of the dynamo to produce the observed large-scale field
has come under increasing scrutiny  
(Cattaneo and Vainshtein 1991, Vainshtein and Rosner 1991,
Kulsrud and Anderson 1992, Brandenburg 1994, Field 1996, 
Blackman 1996, Chandran 1997, Subramanian 1997). 
It appears therefore that understanding large-scale dynamo action 
due to helical turbulence is also closely linked with understanding
the dynamics of the small-scale fluctuating component of
the magnetic field. This forms the prime
motivation for the present work, where we revist the
dynamics of the fluctuating field ab initio.

Fluctuating field dynamics is best studied in terms of 
the dynamics of magnetic correlation functions (see below). Kazantsev (1968)
derived the equations for the longitudinal correlations in homogeneous,
isotropic , Markovian turbulence, without mean helicity, using a 
diagrammatic approach. Vainshtein and Kitchatinov (1986)
incorporated the effects of helicity, and derived equations
for both helical and longitudinal correlations. They did not however
give details of the algebra. We therefore first rederive these equations
here, in a fairly transparent manner. Our work
generalises the Kazantsev equations to helical turbulence,  
and corrects a sign error of Vainshtein and Kitchatinov (1986)
in the helical terms, which could be important for understanding
the back reaction of the small-scale field on the alpha effect (see below).

The above mentioned works
on small-scale field dynamics were also purely kinematic.
As the field grows one expects the back reaction due to the
Lorentz force to become important. The full MHD problem,
involving both the Euler and induction equation, presents
a formidable challenge, which we do not take up here.
However in order to get a first look at the effect of nonlinearities,
we consider a simpler nonlinear modification 
to the velocity field due to the generated magnetic field,
as would obtain for example in a partially ionised medium. 
Note that the gas in galaxies, proto stellar disks and in the universe
just after recombination are all likely to be only partially ionised. 
In such a medium, the Lorentz force on the charged component
will cause a slippage between it and the neutrals.
Its magnitude is determined by the
balance between the Lorentz force and the ion-neutral collisions. 
This drift, called ambipolar drift (Mestel and Spitzer 1956, 
Draine 1986, Zweibel 1988), can be 
incorporated as a field dependent addition to the fluid
velocity in the induction equation. This makes
the induction equation nonlinear and provides us with a
model nonlinear problem to study.  Such a 
modification of the velocity field has also been
used by Pouquet {\it et al.} (1976) and 
Zeldovich {\it et al.} (1983) (pg. 183) to discuss 
nonlinear modifications to the alpha effect.

We incorporate the effects
of this nonlinearity while deriving the evolution 
equations for both the mean field and fluctuating field correlations,
in section 3 and 4. 
The presence of a nonlinear term due to ambipolar
type drift in the induction equation implies that lower moments
couple to higher order moments. 
Some form of closure has to be assumed.
We adopt here a Gaussian closure.
The mean field and the magnetic field correlations then obey a 
set of nonlinear partial
differential equations with the nonlinearity appearing as time-dependent
co-efficients involving the average properties of the fluctuating 
field itself. In the sections which follow we apply these equations to study
a number of astrophysically interesting problems.

We begin in section 5 by examining small-scale dynamo action
in the galactic context. This problem
was studied by KA who looked at the evolution of 
the magnetic energy spectrum in wavenumber space,
taking the limit of large $k$ (small scales).
In contrast to their treatment, the co-ordinate space approach 
adopted here, allows us to implement the boundary conditions for 
the magnetic correlation function, at both large and small-scales.
Using the WKBJ approximation, 
we derive conditions for growth of small-scale fields
in a model Kolmogorov turbulence and study the eigen 
functions for the longitudinal magnetic correlation (see below).
Most earlier work also discussed small scale dynamo action for 
the case when the turbulent velocity 
has a single scale (cf. Kleeorin {\it et al.}
1986, Ruzmaikin {\it et al.} 1989 and references therein),
Our work extends this to the context where multiple scales are present.

The nonlinear effects of ambipolar drift, 
on the small-scale dynamo are considered in section 6.
The results of section 5 and 6, suggests a useful visualisation of the spatial
distribution of the small-scale dynamo generated field 
(cf. Zeldovich et. al, 1983).
We argue that in the galactic context, the magnetic field generated
by small scale dynamo action,
concentrates into thin (perhaps ropy) structures. 
One of the aims of our work here
is to lay a framework for a companion
paper (Subramanian, 1997 : Paper I). In Paper I
we have built upon the results obtained in sections 5 and 6,
to discuss in detail how the small-scale dynamo may saturate in
the galactic context, in a manner which preserves large-scale dynamo
action.
 
Section 7 concentrates on the kinematic evolution of the helical
component of the magnetic correlations. 
The study of helical magnetic correlations has been 
mostly neglected in the literature, since they do
not drastically affect the small-scale dynamo. 
However the evolution of current helicity is 
important in determining
the back reaction of small-scale fields
on the $\alpha$-effect in the standard dynamo equation (cf.
Pouquet {\it et al.} 1976, Zeldovich {\it et al.} 1983, Gruzinov 
and Diamond 1994, Bhattacharjee and Yuan 1995).
Numerical simulations by Tao {\it et al.}, 1993 indicated that
the alpha effect is drastically decreased by the growing
small-scale field (see however Brandenburg 1994).
We will examine this issue in terms of our approach.

In section 8, we study the nonlinear 
effects due to the current helicity. We point out
that the current helicity may increase temporarily 
due to its non linear coupling with the logitudinal magnetic
correlation function. We also comment on the
possibility of a non linear dynamo driven by the 
magnetic alpha effect and on the ordering of small scale
fields due to relaxation and selective decay.
The final section presents a summary of the results.

\section{Mathematical preliminaries }

In a partially ionised medium the magnetic field evolution is 
governed by the induction equation
\begin{equation}
(\partial {\bf B}/ \partial t) =
{\bf \nabla } \times ( {\bf v}_i \times {\bf B} - 
 \eta {\bf \nabla } \times {\bf B}), 
\end{equation} 
where ${\bf B}$ is the magnetic field,  
${\bf v}_i$ the velocity of the ionic component of the fluid and
$\eta$ the ohmic resistivity. The ions experience the Lorentz force due 
to the magnetic field. This will cause them to drift with respect to
the neutral component of the fluid. If the ion-neutral 
collisions are rapid enough, one can assume that the Lorentz force on 
the ions is balanced by their friction with the neutrals. Under this 
approximation the Euler equation for the ions reduces to :
\begin{equation}
\rho_i \nu_{in} ({\bf v}_i -{\bf v}_n ) \equiv \rho_i \nu_{in} {\bf v}_D 
= [({\bf \nabla } \times {\bf B}) \times {\bf B}]/ (4\pi),
\end{equation}
where $\rho_i$ is the mass density of ions, $\nu_{in}$ the ion-neutral 
collision frequency and ${\bf v}_n $ the velocity of the neutral particles.
We have also defined here ${\bf v}_D$, the ambipolar drift velocity.

Using the Euler equation for the ions and 
substituting for ${\bf v}_i$, the induction equation 
becomes the nonlinear equation 
\begin{equation}
{\partial {\bf B}\over \partial t} =
{\bf \nabla } \times \left[ {\bf v}_n \times {\bf B}+ 
a[(({\bf \nabla } \times {\bf B}) \times {\bf B}) \times {\bf B} ]
 -\eta {\bf \nabla } \times {\bf B} \right], 
\label{basic}
\end{equation}
where we have defined
\begin{equation}
a = {1 \over 4 \pi \rho_i \nu_{in} }
\label{adef}
\end{equation}

We have derived the above equation as one which describes
the effect of ambipolar drift. However, one can also view
Eq. (\ref{basic}) as describing 
a model nonlinear problem, where the nonlinear
effects of the Lorentz force are taken into account as
simple modification of the velocity field. 
Such a phenemenological modification of the velocity field has in fact been
used by Pouquet {\it et al.} (1976) and 
Zeldovich {\it et al.} (1983) (pg. 183) to discuss 
nonlinear modifications to the alpha effect.

The velocity field ${\bf v}_n$ 
is taken to be prescribed independent of the magnetic field.
We will assume ${\bf v}_n$ has 
a turbulent stochastic component ${\bf v}_T$ over and above a smooth 
component ${\bf v}_0$, that is ${\bf v}_n = {\bf v}_0 + {\bf v}_T$.
Since ${\bf v}_T$ is stochastic, Eq.\ (\ref{basic}) becomes 
a stochastic partial differential equation. Its solution depends on 
the statistical properties of the velocity field ${\bf v}_T$. 

We assume ${\bf v}_T$ to be an isotropic, homogeneous,
Gaussian random velocity field with 
zero mean. For simplicity, in this work, we also assume ${\bf v}_T$ to have 
a delta function correlation in time (Markovian approximation) and 
its two point correlation to be specified as 
\begin{equation}
<v^i_T({\bf x},t)v^j_T({\bf y},s) > = T^{ij}(r) \delta (t-s) 
\end{equation}
with
\begin{equation}
T^{ij}(r) =
T_{NN}[\delta^{ij} -({r^i r^j \over r^2})] +
 T_{LL}({r^i r^j \over r^2}) +
C\epsilon_{ijf} r^f .
\end{equation}
Here $<>$ denotes averaging over an ensemble of the stochastic velocity
field ${\bf v}_T$, $r = \vert {\bf x} -{\bf y} \vert$, $r^i = x^i -y^i$ and 
we have written $T^{ij}(r)$ in the form appropriate for a statistically
isotropic and homogeneous tensor (cf. Landau $\&$ Lifshitz 1987). 
$T_{LL}(r)$ and $T_{NN}(r)$ are the 
longitudinal and transverse correlation functions for the velocity 
field while $C(r)$ represents 
the helical part of the velocity correllations.
If ${\bf v}_T$ is assumed to be divergence free (which we do here),
then
\begin{equation}
 T_{NN} = {1 \over 2r} {\partial \over \partial r} (r^2 T_{LL}(r)). 
\label{tntl}
\end{equation} 

The stochastic Eq.\ (\ref{basic}) can now be 
converted into equations for the various moments of the 
magnetic field. To derive these equations we proceed as follows:
(see also Zeldovich {\it et al.} 1983, Chapter 8)

\section{ Mean field evolution}

Let the magnetic field at an initial time say $t=0$ be ${\bf B}({\bf x},0)$.
Then, at an infinitesimal time $\delta t$ later, the field 
is given iteratively by
\begin{equation}
{\bf B}({\bf x},\delta t) = {\bf B}({\bf x},0) + 
\delta t \ \eta  {\bf \nabla}^2 {\bf B}({\bf x},0) +
\int_0^{\delta t} dt {\bf \nabla } \times {\bf E}_A({\bf x},t) + 
\int_0^{\delta t}dt \int_0^{\delta t} ds {\bf \nabla } \times 
{\bf E}_B({\bf x},t,s) 
\label{sol}
\end{equation}
where we have defined  
${\bf E}_A({\bf x},t)= {\bf V}({\bf x},t)\times {\bf B}({\bf x},0)$, 
${\bf E}_B({\bf x},t,s) = {\bf v}_T({\bf x},t)\times 
[ {\bf \nabla } \times ( {\bf v}_T({\bf x},s)\times {\bf B}({\bf x},0))] $ and
${\bf V}={\bf v}_n +{\bf v}_D$. 
We have also retained only terms which are potentially of order $\delta t$,
and so will survive in the limit $\delta t \to 0$. Note that the
last term in the above equation is potentially first order
in $\delta t$, because of the presence of the stocahstic turbulent
velocity field ${\bf v}_T$.

The time evolution of the mean magnetic field ${\bf B}_0=<{\bf B}>$ 
can be deduced by taking the ensemble average of Eq.\ (\ref{sol}) .
As before $<>$ denotes averaging over an ensemble of the stochastic velocity
field ${\bf v}_T$. On using the fact that 
${\bf v}_T$ at time $t$ is not correlated with either the initial
magnetic field  ${\bf B}({\bf x},0)$ 
or the initial  perturbed field
$\delta {\bf B}({\bf x},0) = {\bf B} - <{\bf B} >$ 
and taking the limit $\delta t \to 0$ 
we get after some straight forward algebra
\begin{equation}
{\partial {\bf B}_0\over \partial t} =
{\bf \nabla } \times [ {\bf v}_0 \times {\bf B}_0 + 2C(0) {\bf B}_0
 -(\eta  +T_{LL}(0) )
 {\bf \nabla } \times {\bf B}_0 ]  +
< {\bf \nabla} \times 
{\bf v}_D \times {\bf B} > 
\label{mean} 
\end{equation}
The effect of the turbulent velocity is to introduce the standard extra 
terms representing the $\alpha$-effect with 
\begin{equation}
\alpha = 2C(0) = 
-{1 \over 3} \int <{\bf v}_T({\bf x}, t).{\bf \nabla} \times 
{\bf v}_T({\bf x},s) > ds
\end{equation}
 and an extra turbulent contribution to the diffusion 
\begin{equation}
\eta_T = T_{LL}(0) = {1 \over 3} \int <{\bf v}_T({\bf x},t).
{\bf v}_T({\bf x},s)> ds . 
\end{equation}

Over and above these terms the effect of ambipolar drift 
is to introduce an extra EMF represented by the 
last term in Eq.\ (\ref{mean}) . This extra term involves the 
third moment of the magnetic field. Similarly as we will see below 
a consideration of the equation for the magnetic correlation function 
will introduce the fourth moment of the magnetic field. This will 
lead to an infinite hierarchy of equations for the moments which 
can be only truncated by assuming some form of closure.
We assume below, for analytic tractability, 
that the magnetic field fluctuation $\delta {\bf B} =
{\bf B} -{\bf B}_0$ is also a homogeneous, isotropic, 
Gaussian random field with zero mean.
Its equal time two point correlation is given by 
\begin{equation}
<\delta B^i({\bf x},t) \delta B^j({\bf y},t) > = M^{ij}(r,t) = 
M_N[\delta^{ij} -({r^i r^j \over r^2})] + 
M_L ({r^i r^j  \over r^2}) + H \epsilon_{ijf} r^f .
\end{equation}
(Here the averaging is a double ensemble average over both the 
stochastic velocity and stochastic $\delta{\bf B}$ fields, although we
indicate only one angular bracket). 
The functions $M_L(r,t)$ and $M_N(r,t)$ are the longitudinal 
and transverse correlation functions for the magnetic field 
while $H(r,t)$ represents 
the helical part of the correlations.
 Note that $H$ is proportional to the current-field
correlation, rather than the field-vector potential
correlation (see below). 
Since ${\bf \nabla}.{\bf B}=0$, $M_N$ and $M_L$ 
are related by
\begin{equation}
 M_N = {1 \over 2r} {\partial \over \partial r} (r^2 M_L(r)). 
\label{bnbl}
\end{equation} 
The nonlinear term due to ambipolar drift, 
$ < {\bf \nabla} \times {\bf v}_D \times {\bf B} > $ 
in Eq.\ (\ref{mean}) is then given by
\begin{equation}
{\bf \nabla} \times [{2 a \over 3}  <\delta {\bf B} . {\bf \nabla }\times 
\delta {\bf B}> + a ({\bf B}_0 . {\bf \nabla }\times {\bf B}_0) ]{\bf B}_0 
-[{2a \over  3} < \delta {\bf B} . \delta {\bf B}> + a ({\bf B}_0^2) ]
{\bf \nabla} \times {\bf B}_0 .
\label{non}
\end{equation} 
Using the form for the magnetic correlation function we 
have  
\begin{equation}
< \delta {\bf B} . {\bf \nabla }\times 
\delta {\bf B}> = - 6 H(0,t) \ {\rm and} \
 < \delta {\bf B} . \delta {\bf B}> = 3M_L(0,t) . \label{mrel}
\end{equation}
So the mean magnetic field satisfies the equation 
\begin{equation}
{\partial {\bf B}_0\over \partial t} =
{\bf \nabla }\times [ {\bf v}_0 \times {\bf B}_0  
+ \alpha_{eff}
 {\bf B}_0  - \eta_{eff} 
 {\bf \nabla } \times {\bf B}_0 ] .
\label{meanfin}
\end{equation} 
where 
\begin{equation}
\alpha_{eff} = 2C(0) - 4aH(0,t) 
+ a ({\bf B}_0 . {\bf \nabla }\times {\bf B}_0) 
\end{equation}
\begin{equation}
\eta_{eff}= \eta   +T_{LL}(0) + 2aM_L(0,t) + a {\bf B}_0^2
\end{equation}

The effect of ambipolar drift 
(or the field dependent addition to the fluid velocity)
on the evolution of 
the mean field is therefore to modify the $\alpha$-effect and 
the diffusion of the mean field. When one starts from small seed 
fields, the additional nonlinear terms 
depending on the mean field itself are subdominant to the 
terms depending on the fluctuating field for most part
of the evolution; since we will find below that the small
scale fields grow much more rapidly compared to the large-scale 
mean field. When ambipolar drift is taken into account, 
the small-scale fluctuating fields contribute an extra 
diffusion term to the mean field evolution, 
proportional to their energy density.
Also the alpha effect is modified by the addition of a term
proportional to the mean 
current aligned component (or current helicity, $H(0,t)$) 
of the magnetic field fluctuations. Some aspects of mean field
dynamos incorporating ambipolar drift has been discussed
by Zweibel (1988) and Proctor and Zweibel (1992).

\section{Evolution of the correlation tensor of magnetic fluctuations} 

The derivation of these equations involves straightforward 
but rather tedious algebra. We therefore only outline the steps and the 
approximations below leaving out most of the algebraic details.
We start by noting that 
\begin{eqnarray} 
(\partial M_{ij}/\partial t) &=&
(\partial /\partial t)(< \delta B_i({\bf x},t)\delta B_j({\bf y},t) > )
\nonumber\\ &=&
[(\partial / \partial t)(< B_i B_j>) - (\partial / \partial t)(<B_i><B_j> ) 
].
\end{eqnarray} 
The second term in the square brackets is easy to evaluate using the
equation for the mean field. The first term can be evaluated using
Eq.\ (\ref{basic}) and the fact that 
\begin{equation}
(\partial / \partial t)( B_i({\bf x},t) B_j({\bf y},t)) = 
B_i({\bf x},t)(\partial  B_j({\bf y},t)/ \partial t) + 
(\partial  B_i({\bf x},t)/ \partial t)B_j({\bf y},t). 
\label{qeq}
\end{equation}
The resulting equation can again be solved iteratively to get 
an equation for $(\partial M_{ij} / \partial t)$ which depends on the 
the turbulent velocity correlations $T_{ij}$, the mean velocity field
${\bf v}_0$ and the mean magnetic field ${\bf B}_0$ and most importantly 
a non-liner term incorporating the effects of ambipolar drift. We get
\begin{eqnarray}
{\partial M_{ij} \over \partial t}
&=& <\int \ ^yR_{jpq}\left[v_T^p({\bf y},t) \ ^xR_{ilm}(v_T^l({\bf x},s) 
[M_{mq}+ B_0^m({\bf x}) B_0^q({\bf y}))\right ] ds > \nonumber\\
&+&
 <\int \ ^xR_{ipq}\left[v_T^p({\bf x},t) \ ^yR_{jlm}(v_T^l({\bf y},s) 
[M_{qm} +  B_0^q({\bf x}) B_0^m({\bf y}))\right ] ds >  
\nonumber\\ &+&
<\int \ ^yR_{jpq}\left(v_T^p({\bf y},t) \ ^yR_{qlm}(v_T^l({\bf y},s) 
M_{im})\right ) ds >  \nonumber\\ &+&
<\int \ ^xR_{ipq}\left(v_T^p({\bf x},t) \ ^xR_{qlm}(v_T^l({\bf x},s) 
M_{mj})\right ) ds > \nonumber\\ 
&+& \eta [ \nabla_y^2M_{ij} + \nabla_x^2 M_{ij}] + 
^yR_{jpq}\left(v_0^p({\bf y}) M_{iq} \right) +
^xR_{ipq}\left(v_0^p({\bf x}) M_{qj} \right)
\nonumber\\ 
&+& 
^yR_{jpq}\left(<v_D^p({\bf y}) \delta B_i({\bf x})B_q({\bf y})> \right) +
^xR_{ipq}\left(<v_D^p({\bf x}) B_q({\bf x})\delta B_j({\bf y})> \right) 
\label{meq}
\end{eqnarray}
where we have defined the operators 
\begin{equation}
^xR_{ipq}= \epsilon_{ilm}\epsilon_{mpq}
(\partial /\partial x^l) \ {\rm and} \ ^yR_{ipq}= \epsilon_{ilm}\epsilon_{mpq}
(\partial /\partial y^l) .
\end{equation}

The first two terms on the RHS of  Eq.\ (\ref{meq}) 
represent the effect of velocity correlations on 
the magnetic fluctuations ($M_{ij}$) and the mean field ($B_0^i$).
The next two terms the "turbulent transport"
of the magnetic fluctuations by the turbulent velocity, the 5th and
6th terms the "microscopic diffusion". The 7th and 8th terms
the transport of the magnetic fluctuations by the mean velocity.
The last two nonlinear terms give the effects of the back reaction 
due to ambipolar drift on the magnetic fluctuations. 

We note that the effects of the mean velocity and magnetic fields 
are generally subdominant to the effects of the fluctuating
velocity and magnetic fields. First the mean fields vary
in general on a scale much larger than the fluctuating fields.
If one neglects the effect of velocity shear due to
the mean velocity, on the fluctating field, then one
can transform away the mean velocity by going to
a different reference frame. The mean magnetic
field, ofcourse cannot be transformed away. But
we will see that, it grows at a rate
much slower than the fluctuating field. So when one
starts from small seed magnetic fields, for most
part of the evolution, we can neglect its effects.
In what follows we will drop the term involving the
mean fields. Due to the above reasons, we also
continue to treat the statistical properties
of the magnetic fluctuations as being homogeneous and 
isotropic, and use $M_{ij}({\bf x}, {\bf y}, t) = M_{ij}(r,t)$
as before.

All the terms in the above equation, can be further simplified by using the 
properties of the magnetic and velocity correlation functions.
In order to obtain equations for $M_L$ and $H$, we 
multiply Eq.\ (\ref{meq})  by $(r^ir^j )/ r^2$ and $\epsilon_{ijf}r^f$
and use the identities 
\begin{equation}
M_L(r)= M_{ij}(r^ir^j / r^2), \  
H(r) =M_{ij} \epsilon_{ijf}r^f/ (2 r^2) .
\end{equation}
We have given some steps in simplifying the first two terms
in appendix A. The 3rd and 4th terms add to give a contribution 
$4C(0)\epsilon_{jqm}(\partial M_{im}/\partial r^q) + 2T_{LL}(0) 
\nabla^2 M_{ij}$ to the RHS of Eq.\ (\ref{meq}) , hence 
justifying their being called  "turbulent transport" of $M_{ij}$.

The last two nonlinear terms give the effects of the back reaction 
due to ambipolar drift on the magnetic fluctuations. 
In evaluating this term, we 
will neglect the effects of the subdominant mean field compared to 
the back reaction effects of the fluctuating field. In this case the 
nonlinear terms add to give a contribution 
$-8aH(0,t)\epsilon_{jqm}(\partial M_{im}/\partial r^q) + 4aM_L(0,t) 
\nabla^2 M_{ij}$ to the RHS of  Eq.\ (\ref{meq}) . The 
Gaussian assumption of the magnetic correlations results in 
the nonlinearity of this term appearing as a nonlinearity in the 
coefficient, rather than the correlation function itself.
Gathering together all the terms, 
we get for the coupled evolution equations for $M_L$ and $H$ :
\begin{equation}
{\partial M_L \over \partial t} = {2\over r^4}{\partial \over \partial r}
(r^4 \kappa_N {\partial M_L \over \partial r}) + G M_L - 4\alpha_N
H  \label{mleq}
\end{equation}
\begin{equation} {\partial H\over \partial t} = 
{1\over r^4}{\partial \over \partial r}
\left(r^4  {\partial \over \partial r}(2\kappa_N H 
+ \alpha_N M_L)\right) 
\label{mheq}
\end{equation}
where we have defined 
\begin{eqnarray}
&&\kappa_N = \eta + T_{LL}(0) - T_{LL}(r) + 2aM_L(0,t) \nonumber\\ 
&&\alpha_N = 2C(0) - 2C(r) -4aH(0,t) \nonumber\\
&&G = - 4\left[ {d\over dr}({T_{NN}\over r}) + {1\over r^2} {d\over dr}
(rT_{LL})
\right] 
\end{eqnarray}

These equations together with Eq.\ (\ref{meanfin}) for the mean magnetic field
are an important result of this work. They form a closed 
set of nonlinear partial differential
equations for the evolution of both the mean magnetic 
field and the magnetic fluctuations, incorporating the 
back reaction effects of ambipolar drift (or a magnetic field dependent 
addition to the velocity). 
For non-helical turbulence the equation for $M_L$ excluding
nonlinear effects was first derived by Kazantsev (1968).
We note that Eq.\ (\ref{mleq}) and 
Eq.\ (\ref{mheq}) without the inclusion 
of the non linear terms due to ambipolar drift, have been derived in 
a different fashion by Vainshtein and Kichatinov (1986). We get exactly 
their equation (27) for $M_L$ and $H$, except for a sign 
difference in front of the $\alpha_N$ terms. We believe that our equations 
have the correct sign (see also below).

The terms involving $\kappa_N$ in equations \ (\ref{mleq}) and \ (\ref{mheq})
represent the effects of diffusion on the magnetic correlations. 
The diffusion coefficient includes the effects of 
microscopic diffusion ($\eta$) and a scale-dependent 
turbulent diffusion ($T_{LL}(0)-T_{LL}(r)$).
The effect of ambipolar drift, under our approximation of Gaussian
magnetic correlations, is to add to the diffusion
coefficient an amount $ 2aM_L(0,t)$ ; a term proportional 
to the energy density in the fluctuating fields.
Similarly $\alpha_N$ 
represents first a scale dependent $\alpha$-effect ($2C(0) -2C(r)$) and 
the effect of ambipolar drift is to decrease this by $4aH(0,t)$, 
an amount proportional to the mean 
current helicity of the magnetic fluctuations.
Ambipolar drift has therefore very similar effect on 
the magnetic fluctuations as on the mean field.
The addition of these terms makes equations (\ref{mleq}) and (\ref{mheq})
nonlinear, with the non linearity appearing in the coefficients.
The term proportional to $G(r)$, allows for 
for the rapid generation of magnetic fluctuations by velocity shear 
and the existence of a small-scale dynamo
{\it independent} of the large-scale field (cf. also KA, 
Vainshtein and Kichatinov 1986). 

An important facet of MHD equations is the conservation of
magnetic helicity, in the absense of microscopic diffusion.
Magnetic helicity is defined as
\begin{equation}
I_M = \int {\bf A}.{\bf B} \ d^3{\bf x}
\end{equation}
where ${\bf A}$ is the vector potential. One can show, (cf. Moffat 1978)
$I_M$ is conserved in the
ideal limit. In case we have no mean fields, we also have 
$I_M = \cal V <{\bf A}.{\bf B}>$ and 
\begin{equation}
{1 \over \cal V} {dI_M \over dt} = -2\eta < {\bf B}.({\bf \nabla} \times 
{\bf B}) >.
\label{imcon}
\end{equation} 
(Note that we are
assuming that integration over a large volume $\cal V$ 
is same as the ensemble average).
It is interesting to display this conservation
explicitely from our equations. This will also act as check on some part
of the algebra. 

For this let us go over to the vector
potential representation of the fluctuating field and
write $\delta {\bf B} = {\bf \nabla} \times \delta {\bf A}$. We also
define the equal time correlation fuction of the fluctuating component of
the vector potential as $P^{ij} = <\delta A^i \delta A^j >$.
Since we have taken the fluctuting magnetic field as a
homogeneous, isotropic, Gaussian 
random field, $P_{ij}$ will also satisfy this property.
So one can write in general
\begin{equation}
P^{ij}(r,t) = P_N[\delta^{ij} -({r^i r^j \over r^2})] + 
P_L ({r^i r^j  \over r^2}) + P_H \epsilon_{ijf} r^f .
\end{equation}
Here the functions $P_L(r,t)$ and $P_N(r,t)$ are the longitudinal 
and transverse correlation functions for the vector potential. 
$P_H(r,t)$ represents 
the helical part of the correlations and is related to the
magnetic helicity of the fluctuating field, 
with 
\begin{equation}
P_H(0,t) = -< \delta {\bf A} . \delta {\bf B }>/6 .
\label{pho}
\end{equation}
So to determine the evolution of the average
magnetic helicity of the fluctuating field, 
we have to evaluate the time evolution of $P_H(0,t)$. 

One can easily relate correlation function $H$,
representing the current-field correlation,
to $P_H$, by using the definitions of the two quantities. We have
\begin{equation}
H(r,t) = -{ r^f \over 2r^2} \epsilon_{ijf}\epsilon_{ilm} 
\epsilon_{jpq} P_{mq,lp} = - {1\over r^4}{\partial \over \partial r}
\left(r^4  {\partial P_H \over \partial r} \right)
\label{hph}
\end{equation}
Substituting this relation in the evolution equation (\ref{mheq}) 
for $H$, we get
\begin{equation} 
{1\over r^4}{\partial \over \partial r} \left[r^4  
{\partial  \over \partial r} \left(
{\partial P_H \over \partial t}  + 2\kappa_N H + \alpha_N M_L \right) \right]
= 0.
\label{pheqi}
\end{equation}
Since all the correlations die off at spatial infinity the integral
of this equation then gives,
\begin{equation}
{\partial P_H(r,t)\over \partial t} = -  (2\kappa_N H + \alpha_N M_L) 
\label{pheq}
\end{equation}

Now suppose we look at the evolution of the mean helicity
of the fluctuating fields, in the absense of the mean
magnetic field. Then conservation of $I_M$ implies
conservation of $P_H(0,t)$. 
To see if this conservation 
obtains, take the $r \to 0$ limit of the Eq. (\ref{pheq}).
We get 
\begin{equation}
{\partial P_H(0,t) \over \partial t} = -2\eta H(0,t)
\label{phcon}
\end{equation}
So when $\eta=0$, the above equation shows that $P_H(0,t) = $ constant,
independent of time, as required. Also when $\eta \ne 0$, and
the mean fields are zero, then 
Eq. (\ref{phcon}) is identical to Eq. (\ref{imcon}) , as
required. It is heartening to
note that the nonlinear equation that we have derived 
for the helical part of the magnetic correlations,
incorporating the effects of ambipolar type drifts, does indeed
embody magnetic helicity conservation in the appropriate
limit, as required.

Note that if the signs in front of the $\alpha_N$
term had been different from the one we get, then the 
ambipolar drift terms in the equation for $P_H$ would
not have cancelled out when we take the $r\to 0$ limit in
Eq. (\ref{pheq}).
And we could not recover magnetic helicity conservation
as above. This provides another check that we have indeed got
the relative sign of these terms right. Also from the
same argument, we note that
one cannot have non-zero additions to the alpha
term in the $r\to 0$ limit (the $-4aH(0,t)$ term in $\alpha_N$) 
without a corresponding addition to the diffusion term 
(the $2aM_L(0,t)$ term in $\kappa_N$). 

We now consider §me of the implications of these equations
for the evolution of small-scale fields.

\section{ Kinematic evolution of $M_L$ and the small-scale dynamo}

We begin by first studying the kinematic evolution of $M_L$,
ignoring the effects of the nonlinear coupling terms
due to ambipolar drift. In particular we consider the 
small-scale dynamo in galactic turbulence with a model Kolmogorov spectrum.
The results obtained in this section will be used to 
set the framework for Paper I. It also extends the analysis of
the small-scale dynamo, due to turbulence with
a single scale (cf. Zeldovich {\it et al.} (1983), 
Kleeorin {\it et al.} (1986)) to the situation when a
range of scales are present.

First we note that in many contexts the coupling term 
$\alpha_N H$, due to helicity fluctuations of the
velocity, has negligible influence on the evolution of $M_L$. 
A canonical estimate in the galactic context (see Zeldovich et al. 1983) 
is $\alpha_N \sim 2C(0)  \sim (V^2\tau /h)(\Omega\tau) \sim \Omega \tau V(L/h)$
and $\kappa_N \sim T_{LL}(0) \sim V^2\tau \sim VL$. Here
where $h$ is the disk scale height, $\Omega$ the rotation frequency,
$V$, $\tau$ and $L$ are the velocity, correlation time, and correlation 
lengths for the energy 
carrying eddies of the turbulence. Assuming $H\sim M_L/h$,
the importance of the 
coupling term $\alpha_N H$ in Eq. (\ref{mleq}) to the other 
terms is $\sim (\alpha_NH/(\kappa_N M_L/L^2)) \sim
\Omega\tau (L/h)^2 << 1$ in general. (Even for $\tau \sim L/V$ we generally 
have $\Omega\tau <<1$). So it is an excellent approximation 
to neglect the coupling to $H$ in examining evolution of $M_L$. We will refine 
this estimate in the next section after studying the evolution of $H$ and
see that the above approximation is even better.
 
The evolution for $M_L$ 
can then be transformed into a Schrodinger-type equation 
by defining $\Psi =r^2\sqrt{\kappa_N}M_L$. We get
\begin{equation}
{1\over 2}{\partial \Psi \over \partial t} = 
\kappa_N{\partial^2 \Psi \over \partial r^2} - U(r,t)\Psi \label{sievol}
\end{equation}
where for a divergence free velocity field, the "potential"
\begin{equation}U(r,t) =  T_{LL}^{\prime\prime} + {2 \over r} T_{LL}^{\prime} 
+ {\kappa_N^{\prime\prime} \over 2}
-{ (\kappa_N^{\prime})^2 \over 4\kappa_N} + {2\kappa_N \over r^2} .
\label{pot}
\end{equation}
The boundary condition on $M_L(r,t)$ is that 
it be regular at the origin and that $M_L(r,t)\to 0$ as $r\to \infty$.

In the kinematic limit, note that $\kappa_N$, $T_{LL}$ and hence
the potential $U$ are time independent. 
Equation (\ref{sievol}) then admits eigenmode solutions
of the form $\Psi(r,t) = \exp(2\Gamma t)\Phi(r)$ where
\begin{equation}
\kappa_N (d^2 \Phi/dr^2) -( \Gamma + U)\Phi = 0 . 
\label{phievol}
\end{equation} 
So there exists a possibility of growing modes with $\Gamma > 0$, 
if one can have $U $ sufficiently negative 
in some range of $r$. The problem of 
having a small-scale dynamo and rapid growth of magnetic fluctuations,
reduces to that of having bound states 
in the potential $U$. 

To see what this requires let us consider a
model problem where the behaviour $T_{LL}(r)$ simulates Kolmogorov  
turbulence (Vainshtein 1982);
\begin{eqnarray} 
 T_{LL}(r) &=& {VL\over 3} 
 [1 - R_e^{1/2}({r\over L})^2] \quad {\rm for} \ 0 < r < l_c \nonumber\\
&=& {VL\over 3}[1 - ({r\over L})^{4/3}] \quad
{\rm for} \ l_c < r < L \nonumber\\
&=& 0 \qquad {\rm for} \ r > L  
\label{tll}
\end{eqnarray}

Here $l_c \approx L R_e^{-3/4}$ is the cut off scale of the
turbulence, where $R_e = VL/ \nu$ is the
fluid Reynolds number and $\nu$ is the kinematic viscosity.
For Kolmogorov turbulence, the eddy
velocity at any scale $l$, is $v_l \propto l^{1/3}$,
in the inertial range. So the scale dependent diffusion
coefficient scales as $v_l l \propto l^{4/3}$.
This scaling, also referred to as Richardsan's law,
is the motivation for the form of the
scaling of $T_{LL}$ with $r$ which we have adopted 
for the inertial range of the turbulence.
Note that the structure function  $T_{LL}(r)$ must satisfy the 
condition $T_{LL}^{\prime}(0) = 0$, at the origin. So
for scales smaller than $l_c$ we have continued $T_{LL}$ from 
its value at $r=l_c$ to zero, satisfying this constraint. 
(One can adopt smoother continuations
for $T_{LL}(r)$ at $r=l_c$ and $r=L$, in order to make the potential $U$ 
continuous at these points. But this has little
effect on the conclusions below, since $\Phi$ is determined
by integrals over $U$). 

The potential is then $U=(2\eta /r^2)$ as $r\to 0$ and 
$U= 2(\eta + \eta_T)/r^2$ for $ r > L$.
For $l_c < r < L$ we have  
\begin{equation}
U = {V \over 3 L}\left[-{8 \over 9}({r \over L})^{-2/3} -{(4/9)(r/L)^{2/3}
\over (3/R_m + (r/L)^{4/3})}
+{6 \over R_m}({L^2 \over r^2}) \right] ,
\label{potin}
\end{equation} 
where $R_m = (VL/\eta)$ is the magnetic reynolds 
number at the outer scale of the turbulence.
The Spitzer value of the resistivity gives
$\eta = 10^7 (T/10^4K)^{-3/2} cm^2 s^{-1}$. 
For numerical estimates we generally
take $V = 10 km s^{-1}$ and $L = 100 pc$. For these
turbulence parameters $R_m = 3 \times 10^{19}$. 

The value of the potential at $r=L$ is
$U \approx (V/L)[(2/R_m) -(4/9)]$. 
If $R_m < 9/2$ one can easily see from the above 
expressions that $U$ remains positive for all $r$.
Note that for a bound state to obtain and for the small-scale fields to 
grow, one must have $U$ negative for some range of $r$. 
So a neccessary condition for the small 
scale fields to grow is $R_m >> 1$. In fact the exact critical
MRN, say $R_c$, for growth has to be determined 
numerically but one typically gets $R_c \sim 60$ (cf., Zeldovich et al 1983,
Novikov et al. 1983; see below for a WKBJ derivation of this limit)

Since the velocity at any scale $l$, say $v_l \propto l^{1/3}$ for
Kolmogorov turbulence, the MRN associated
with eddies of scale $l$, is $R_m(l) = v_l l/\eta = R_m (l/L)^{4/3}$.
Using this one can also rewrite the potential $U$ as 
\begin{equation}
U = {v_l \over 3 l}\left[-{8 \over 9}({r \over l})^{-2/3} -{(4/9)(r/l)^{2/3}
\over (3/R_m(l) + (r/l)^{4/3})}
+{6 \over R_m(l)}({L^2 \over r^2}) \right] .
\label{potinl}
\end{equation}
This is exactly of the same form as Eq. \ (\ref{potin})  except that
$L, V$ and $R_m$ are replaced by $l, v_l$ and $R_m(l)$ respectively.
So a number of conclusions about the generation of small-scale fields
can be scaled to apply to an arbitrary scale, $l$, provided we use
the corresponding velocity scale $v_l$ and Reynolds number $R_m(l)$
appropriate to the scale $l$. For example, the condition 
for excitation of small-scale dynamo modes which are concentrated 
at a scale $l$, is also $R_m(l) = R_c >> 1$. Of course,
exact scalability of the results only obtains well in the inertial range,
far away from $l_c$ and $L$.   
The MRN associated with eddies at the 
cut off scale is $R_m(l_c) = v_c l_c/\eta = 
V(l_c/L)^{1/3} L (l_c/L)/\eta = R_m/R_e$. (Here $v_c$ is the
eddy velocity at the cut-off scale).
So if  $R_m/R_e >> 1$, a potential well with $U$ negative, 
extends up to the cut off scale of the turbulence, and these modes
will also grow. 

We will be considering a largely
neutral galactic gas and for this $\nu$ is dominated by 
the neutral contribution.
We take the neutral-neutral collision
to be dominated by H-H collisons with a cross section 
$\sigma_{H-H} \sim 10^{-16} cm^{-2}$, leading to a
kinematic viscosity $\nu \sim v_{th} (1 / n_H \sigma_{H-H})$. 
For a thermal velocity $v_{th} \sim 10 km s^{-1}$ and a
neutral hydrogen number density $n_H \sim 1 cm^{-3}$, as 
say appropriate for a  young galaxy, we have
$\nu \sim 10^{22} cm^2 s^{-1}$, and 
\begin{equation}
R_e = {VL \over \nu } \approx 3 \times 10^{4} V_{10} L_{100}
\label{rey}
\end{equation}
where $V_{10} = (V/ 10 {\rm km s}^{-1})$ and $L_{100} =
(L/ 100 {\rm pc})$. So the condition $R_m/R_e > R_c $ in fact holds in 
the galactic
context. 

It should also be noted that the value of the potential at any $l$, 
$U(l) \sim v_l/l$, is the inverse of the turnover time
of the eddies of scale $l$. The depth of the 
potential well at some scale $l$ reflects the growth 
rate of modes concentrated around that $l$.
And the growth rate of a mode 
extending up to $r \sim l$, say $\Gamma_l \sim U(l) \sim v_l/l \sim l^{-2/3}$,
decreases with increasing $l$.
So when $R_m(l_c) = R_m/R_e > R_c$, 
the small-scale fields tangled at the cut off scale
grow more rapidly than any of the larger scale modes.
These results have also been found in a 
different fashion by KA. To illustrate some of 
these points in a more quantitative fashion we have given a detailed
a WKBJ analysis of Eq.\ (\ref{phievol}) in appendix B.
Below we summarise the main results of this analysis:

\begin{itemize}

\item The WKBJ analysis finds 
a critical value of the MRN, $R_m = R_c \approx 60$, for
the excitation of the small-scale dynamo. Above this critical MRN
the small-scale dynamo can lead to an exponential growth of the
fluctuating field correlated on a scale $L$. We refer
to the eigenmode which is excited for $R_m=R_c$ as the marginal mode.
Further, the equations determining
$R_c$ are the same if we replace $(L,R_m)$ by $(l,R_m(l))$. Therefore,
the critical MRN for excitation of a
mode concentrated around $r\sim l$ is also $R_m(l) = R_c$, as
expected from the scale invariance in the inertial range.

\end{itemize}
 
In the galactic context $R_m >> R_c$; in fact, one also has
$R_m(l_c) = R_m/R_e >> 1$. Hence, small-scale dynamo action
excites modes correlated on all scales from the cut-off
scale $l_c$ to the external scale $L$ of the turbulence.

\begin{itemize}

\item As expected, 
due to small-scale dynamo action, the fluctuating field, 
tangled on a scale $l$, grows exponentially on the corresponding
eddy turnover time scale, 
with a growth rate $\Gamma_l \sim v_l/l \propto l^{-2/3}$. 
In the galactic context, with
$R_m(l_c) = R_m/R_e >> R_c$, 
the small-scale fields tangled at the cutoff scale
grow more rapidly than any of the large-scale modes.

\item The WKBJ analysis gives a growth rate
$\Gamma_c = (v_c/l_c) [5/4 - c_0 ({\rm ln}(R_m/R_e))^{-2} ]$
with $c_0 = \pi^2/12$ for the fastest growing mode. Note that this
is only weakly (logarithmically) dependent on $R_m$, 
provided $R_m$ is large enough.

\end{itemize}

To examine the spatial structure for various 
eigenmodes of the small-scale dynamo, 
it is more instructive to consider the function $w(r,t) = 
<\delta {\bf B}({\bf x},t) .\delta {\bf B}({\bf y},t)> $, 
which measures the correlated 
dot product of the fluctating field ($w(0) = 
<\delta {\bf B}^2 >$ ).   
Firstly there is a general constraint that can be placed
on $w(r)$. Since the fluctuating field is 
divergence free, we have
\begin{equation}
w(r,t) = {1\over r^2} {d\over dr}\left [ r^3 M_L \right ] ,
\label{wr}
\end{equation}
so 
\begin{equation}
\int_0^{\infty} w(r) r^2 dr = \int_0^{\infty} 
{d\over dr}\left [ r^3 M_L \right ]  = 0 , 
\label{consw}
\end{equation}
 since $M_L$ is regular at the origin and vanishes faster than
$r^{-3}$ as $r \to \infty$. Therefore
the curve $r^2w(r)$ should have zero area
under it. Since $w(0) = < (\delta{\bf B})^2>$, 
$w$ is positive near the origin. And
the fluctuating field points in the same direction for small
separation. As one goes to larger values of $r$, there must then 
values of $r$, say $r \sim d$, where $w(r)$ becomes negative.
For such values of $r$, the field at the origin and at a separation
$d$ are pointing in opposite directions on the average. 
This can be intepreted 
as indicating that the field lines, 
on the average are curved on the scale $d$.

\begin{itemize}

\item We find that, in the case $R_m/R_e >> 1$, $w(r)$ is strongly peaked 
within a region $r =r_d \approx l_c (R_m/R_e)^{-1/2}$ about the origin, 
for all the modes. Note that $r_d$ is the diffusive scale 
satisfying the condition $\eta/r_d^2 \sim v_c/l_c$. 
For the most rapidly growing mode, $w(r)$ 
changes sign across $r \sim l_c$ and rapidly decays
with increasing $r/l_c$. For slower growing modes, with 
$\Gamma_l \sim v_l/l$, $w(r)$ extends up to $r \sim l$
after which it decays exponentially. 

\item For the marginal mode
$w(r)$ peaks within a radius $r \sim L /R_c^{3/4} $, changes sign to become
negative at $r \sim L$ and dies rapidly for larger $r/L$.

\end{itemize}

We should point out that a detailed 
analysis of the eigenfunctions can be found in Kleeorin et. al. (1986), 
for the simple case when the longitudinal
velocity correlation function has only a single scale. Their analysis
is also applicable to the mode near the cut-off scale of 
Kolmogorov type turbulence. These authors also give a pictorial 
intepretation of the correlation function, in terms of the
Zeldovich rope-dynamo (cf. Zeldovich {\it et al.} 1983). 
If one adopts this intepretation, the small-scale field can be 
thought as being concentrated in rope like structures with thickness $r_d$ and
curved on a scale upto $\sim l$ for a mode extending to $r\sim l$.
In Paper I, we also elaborate on a qualitative picture 
of the mechanism for the dynamo growth of small scale fields,
and the generation of ropy fields from general initial
conditions.

As the small sclale fields grow
the back reaction due to ambipolar drift will become important. 
We now turn to the nonlinear effects on the small
scale dynamo to see when this can lead to
small-scale dynamo saturation.

\section{ Non-linear effects on the small-scale dynamo}

As the small-scale fields grow, ambipolar drift
adds to the diffusion
coefficient an amount $ 2aM_L(0,t)$ ; a term proportional 
to the energy density in the fluctuating fields.
Similarly, ambipolar drift leads to a decrease of 
$\alpha_N$ by $4aH(0,t)$, 
an amount proportional to the mean 
current helicity of the magnetic fluctuations. 
When considering the evolution of
the longitudinal correlation function, 
we once again neglect the subdominant effect of the 
the coupling term 
$\alpha_N H$, due helicity fluctuations (see above).

As $M_L(0,t)$ grows, its effect then is simply to change $\eta$ to an
effective 
\begin{equation}
\eta_{ambi} = \eta + 2aM_L(0,t) 
\end{equation} 
in the expression for the potential $U(r,t)$. 
One can define an effective MRN, for
fluid motion on any scale of the turbulence
\begin{equation}
R_{ambi}(l) = {v_l l \over \eta_{ambi}} \approx {v_l l \over 2aM_L(0,t) }
\label{rmeffi}
\end{equation}
where $v_l = (l/L)^{1/3} V$ as before.

As the energy density in the fluctuating field 
increases $ R_{ambi}(l)$ decreases.
Firstly, this makes it easier for the field energy to reach the
diffusive scales $r_d \sim l_c/R_{ambi}^{1/2}(l_c)$, from a general 
initial configuration. More importantly 
as this happens the potential well disappears, 
first at small-scales and then progressively at larger and larger scales.
This means that $M_L$ will grow slower and slower.
The detailed evolution of $M_L$ will be complicated. 
However, the decrease of $R_{ambi}$ suggests one possible 
nonlinear saturation mechanism for the small-scale field.
The possibility that the system finds the stationary 
state with $(\partial M_L/\partial t) =0$. 
In such a state, $M_L$ is independent of time.
So, the condition on the critical MRN for the stationary
state to be reached, will
be identical to that obtained in the kinematic stage.
From the discussion on the kinematic evolution of $M_L$, 
the stationary state is an eigenmode for the system which obtains
when the energy density of magnetic fluctuations has grown such that
\begin{equation}
R_{ambi}(L) = {VL\over \eta + 2aM_L(0,t) } = R_c .
\label{rmeff}
\end{equation}
So if $R_{ambi}(L)$ decreases to a value $R_c \sim 60$, 
dynamo action will stop completely. 

Let us consider now whether this condition can obtain
in turbulent, partially
ionised galactic gas. Take for example 
$V\sim 10 {\rm km} \ {\rm s}^{-1}$ and
$L\sim 100 {\rm pc}$ appropriate for galactic turbulence.
Also let us assume that the galaxy had
very nearly primordial composition in its early stage of evolution:
then the ions are mostly just protons and the neutrals are
mostly hydrogen atoms. We estimate in Paper I
that $\rho_i \nu_{in} = n_i \rho_n < \sigma v>_{eff}$ with
$<\sigma v>_{eff} \sim 4 \times 10^{-9} cm^{3} s^{-1}$.
Here $n_i$ is the ion number density.
We then have 
\begin{equation}
R_{ambi}(l)  = {1 \over f(l) } {3\rho_i \nu_{in} l \over 2\rho_n v_l} =
{Q(l) \over f(l)} ,
\label{rambi}
\end{equation}
where $f(l) = B_l^2/(4\pi \rho_n v_l^2)$ is the ratio of the local magnetic
energy density of a flux rope curved on scale $l$, 
to the turbulent energy density $ \rho_n v_l^2/2$
associated with eddies of scale $l$.
Using the value of $\nu_{in}$ as determined
above and putting in numerical
values we get 
\begin{equation}
Q(l) = {3\rho_i \nu_{in} l \over 2\rho_n v_l} 
\sim 1.8 \times 10^{4} n_{-2}
({l \over L})^{2/3} L_{100} V_{10}^{-1}  
\label{rambii}
\end{equation}
where $n_{-2} = (n_i / 10^{-2} cm^{-3}) $, and 
we have assumed a 
Kolmogorov scaling for the turbulent velocity
fluctuations.

One can see from Eq.\ (\ref{rambi}) - (\ref{rambii}) 
that, for typical parameters
associated with galactic turbulence, the MRN incorporating
ambipolar drift is likely to remain much larger than $R_c$ for
most scales of the turbulence, even when the field energy density
becomes comparable to the equipartition value.
Only if the galactic gas is very weakly ionised, 
 with $n_i < 10^{-5.5} {\rm cm}^{-3}$, can ambipolar 
drift by itself lead to small-scale dynamo saturation.
Such small ion densities may perhaps
obtain in the first collapsed objects in the universe, which 
collapse at relatively high redshifts (cf. Tegmark {\it et al.} 1997).
However for most regions of a disk galaxy, the ion density 
is likely to be larger. 
So one expects the field to continue to grow rapidly,
even taking into account ambipolar drift.
Note also that the growth rates for the small-scale 
dynamo generally depends only weakly
on the MRN, provided the MRN is much larger than $R_c$.
(see section 5, and Kleeorin et al. 1986). Therefore, 
we still expect the small-scale dynamo-generated field to grow 
almost exponentially on the eddy turn around time scale,
as long as $R_{ambi} >> R_c$.
The spatial structure of the fluctuating field 
will also remain ropy, as in the kinematic regime, 
as long as $R_{ambi} >> R_c$.

How then does the small-scale dynamo saturate in galaxies?
In Paper I we consider this question in some detail by examining
other nonlinear feedback processes which
could limit the growth of the galactic small-scale dynamo,
taking into account also ambipolar drift.
We briefly suumarise our findings here for the sake
of completeness. The reader is referred to Paper I for details.

It turns out that the effect of the growing
magnetic tension associated with the small-scale
dynamo generated field is crucial. This tension acts to
straighten out the curved flux ropes, while
frictional drag damps the magnetic energy
associated with the wrinkle in the rope. Also, small-scale flux loops
can collapse and disappear. For a significantly neutral geas,
these non-local effects turn out to operate on
the eddy turnover time scale, when the peak field in a flux rope
has grown to a few times the equipartition value. Their net effect is to make
the random stretching needed for the small-scale dynamo
inefficient and hence saturate the small-scale dynamo.
However, the average energy density in the saturated small-scale field is
sub equipartition, since it does not fill the volume.
It is probable then that the small-scale dynamo generated fields
do not drain significant energy from the turbulence, 
nor convert eddy motion of the turbulence
on the outer scale to wave-like motion. So the
diffusive effects needed for the large-scale dynamo
operation can be preserved.
This picture of small-scale dynamo saturation obtains
only when the ion density is less than a
critical value of $n_i^c \sim 0.06 - 0.5 cm^{-3} (n_n/ cm^{-3})^{2/3}$.

We note in passing that in the case of nearly neutral disks
around protostars, the parameters could indeed be such that
small-scale dynamo action saturates due to purely
ambipolar drift (cf. Proctor and Zweibel 1992). 
We now turn to the the evolution of the
current-field correlations.

\section{Kinematic evolution of current helicity and 
$\alpha$-effect suppression }

The evolution of $H$ is governed by Eq.\ (\ref{mheq}) derived in section 4.
We consider first the kinematic linear evolution 
when the nonlinear terms in $\kappa_N$ and $\alpha_N$ are ignored.
We saw above that the coupling of $H$ to $M_L$ is unimportant 
for the evolution of the longitudinal correlation function $M_L$,
which could grow due purely to self-coupling terms.
However for the evolution of $H$, its coupling to $M_L$
provided by the $\alpha_N$ term cannot
be neglected, because without the forcing by the $M_L$ 
term in Eq.\ (\ref{mheq}), 
any initial distribution of $H$ decays with time due to diffusion. 
To see this multiply \ (\ref{mheq}) by $2Hr^4\kappa_N$ and integrate over 
all $r$, neglecting the $\alpha_N M_L$
coupling term. We get  
\begin{equation} {\partial \over \partial t}\left[\int_0^\infty H^2 \kappa_N 
r^4 dr \right] = 
2 \kappa_N H r^4  {\partial \over \partial r}(2\kappa_N H )\vert^\infty_0
- \int_0^\infty dr r^4  [{\partial \over \partial r}(2\kappa_N H )]^2 .
\label{mhcon}
\end{equation}
Neglecting the boundary term for a sufficiently rapidly falling $H$ 
we see that any smooth $H$ will decay with time in the absence 
of the coupling to $M_L$.

So we have to consider the full inhomogeneous partial 
differential equation \ (\ref{mheq}) for $H$, taking $M_L$ to be 
a given function of $r$ and $t$ determined by the analysis
of section 5. The solution to \ (\ref{mheq})\ can be written then 
as the sum of the solution to the homogeneous equation $H_0$ and a 
particular solution, say $H_p$, forced by the presence of $M_L$.
$H_0$ decays with time, as shown above 
and therefore the asymptotic evolution of $H$ in time will be 
governed by $H_p$. To determine $H_p(r,t)$ we use the Green's function
technique (cf. Burton 1989). 
We are particularly interested in determining $H_p$ when 
the longitudinal correlation function $M_L = \exp(\Gamma_n t) M_n(r)$,
where the function $M_n(r)$ is an eigenmode of section 5, with 
growth rate $\Gamma_n$. In this case we can write $H_p(r,t) = H_n(r)
\exp(\Gamma_n t) $, with $H_n(r)$ satisfying the time independent 
differential equation
\begin{equation} \Gamma_n H_n(r) = 
{1\over r^4}{d\over dr}
\left(r^4  {d \over  dr}(2\kappa_N H_n 
+ \alpha_N M_L)\right) \label{mheqr}
\end{equation}
The boundary conditions are that $H_n(r)$ is regular at $r=0$ and
$H_n(r) \to 0$ as $r\to \infty$, as before. 

For general $\kappa_N(r)$ and $\alpha_N(r)$ there are no 
general solutions to \ (\ref{mheqr}). We obtain below approximate
solution using the WKBJ approximation.
To implement the boundary condition at $r=0$, under the WKBJ approximation, 
again it is better to transform to a new radial co-ordinate $x$, 
where $r=e^x$. 
Substituting $H_n(x) = f(x)\exp(-3x/2)/\kappa_N$, equation \ (\ref{mheqr})
then becomes
\begin{equation} 
{d^2f\over dx^2} - \left( {9\over 4} + {\Gamma_n e^{2x} \over 2\kappa_N}
\right) f = - e^{3x/2} \left ({d^2 (\alpha_N M_n)\over dx^2} + 
3{d(\alpha_N M_n)\over dx} \right) 
\label{feq}
\end{equation}
The boundary conditions now become $f(x) \to 0$ as $x \to \pm \infty$.

The WKBJ solution of the homogeneous equation for $f=f_H$ are 
\begin{equation}f_H = F_{\pm}(x) = 
\left( {9\over 4} + {\Gamma_n e^{2x} \over 2\kappa_N }\right)^{-1/4}
\exp (\pm \int_a^x 
 \left( {9\over 4} + {\Gamma_n e^{2x^{\prime}} \over 2\kappa_N(x^{\prime}) }
\right)^{1/2} dx^{\prime})  
\label{wkbF}
\end{equation}
Note that $F_+(x)$ satisfies the left boundary condition $F_+ \to 0$ as 
$r=e^x\to 0$, while $F_-$ staisfies the right boundary condition on $f$.
The Wronskian of the two solutions is $W = F_+d(F_-)/dx - d(F_+)/dx F_- = -2$.
The particular solution to \ (\ref{feq})\ , which satisfies the given 
boundary conditions can then be written as
\begin{equation}f(x) = \int_{-\infty}^{\infty}
dy G(x,y) e^{3y/2} \left ({d^2 (\alpha_N M_n)\over dy^2} + 
3{d(\alpha_N M_n\over dy} \right) \label{fpar}
\end{equation}
where the Green function $G(x,y)$ is given by
\begin{equation}  
G(x,y) = -{1\over W} \left[ \theta(y-x)F_+(x)F_-(y) + 
\theta(x-y)F_+(y)F_-(x)\right].
\label{green}
\end{equation}
Here $\theta(x)$ is the standard heavyside function, which is zero for
$x < 0$ and is unity for positive $x$.

Transforming back to the original radial co-ordinate $r$ we finally have
\begin{equation}
 H_p(r,t) = {\exp(\Gamma_n t) \over 2 \kappa_N(r)} 
\int_0^\infty du ({u\over r})^{3/2} 
{1\over 2} \left[ \theta(u-r) K(r;u) + \theta(r-u) K(u;r)\right]  
\times {1\over u^3}{d\over du}
\left(u^4  {d \over  du}(\alpha_N M_n(u))\right)  
\label{hpsol}
\end{equation}
where we have defined
\begin{equation}
K(r;u) =
\left( {9\over 4} + {\Gamma_n r^2 \over 2\kappa_N(r) }\right)^{-1/4}
\left( {9\over 4} + {\Gamma_n u^2 \over 2\kappa_N(u) }
\right)^{-1/4} 
\exp \left[ \int_u^r 
 \left( {9\over 4} + {\Gamma_n r^{\prime 2}\over 2\kappa_N(r^{\prime}) }
\right)^{1/2} { dr^{\prime} \over r^{\prime}}\right] . 
\label{kdef}
\end{equation}

Of primary interest is the behaviour of $H_P(0,t)$, the time evolution of
the average current helicity associated
with the fluctuating field, since it is this quantity which alters the 
alpha effect in the equation for the mean field evolution. This can be 
evaluated by putting $r=0$ in equation \ (\ref{hpsol})\ . We get  
\begin{equation}  
H_p(0,t) = {\exp(\Gamma_n t) \over \sqrt{6}} 
\int_0^\infty dy  
{ \exp ( - \int_0^y { d y^{\prime} \over y^{\prime}}
 \left[( {9\over 4} + {\kappa_N(0)y^{\prime 2}\over 
\kappa_N(y^{\prime}) })^{1/2}
- {3\over 2} \right] \over 
( {9\over 4} + {\kappa_N(0) y^2\over \kappa_N(y) })^{1/4} }  
\times {1\over y^3}{d\over dy}
\left(y^4  {d \over  dy}({\alpha_N M_n(y) \over 2\kappa_N(0)})\right) 
\label{hpsolo}
\end{equation}
where we have used the identity 
$(u/r)^{3/2} = \exp(\int_r^u (3dx /2x))$ and also 
defined new variables $y =u/a$ and $y^{\prime} = r^{\prime}/a$
with $a^2 = 2\kappa_N(0)/\Gamma_n$.  

We can calculate $H_P(0,t)$ explicitely once the functional forms
of $\alpha_N(y)$, $\kappa_N(y)$ and $M_n(y)$ are specified. 
For this let us consider the 
example of the previous section, 
where we $T_{LL}(r)$ is given by \ (\ref{tll})\ ,
simulating Kolmogorov turbulence. In the previous section
we saw that in the case $R_m/R_e >> 1$, all the modes
were strongly peaked about a radius $r = r_d \sim l_c (R_m/R_e)^{-1/2} $.
And the growth rate $\Gamma_n \sim v_l/l$ for a mode
extending up to $r \sim l$ with $\Gamma_n \sim R_e^{1/2} V/L \equiv \Gamma_c$
for the fastest growing mode.
As a model we therefore adopt a Gaussian form for $M_n(r)$ 
\begin{equation}
M_n(r) = M_n(0) \exp[- {r^2 \over 2 r_d^2} ] =M_n(0) \exp(-b y^2) 
\label{mnex}
\end{equation}
where $b = (a^2/2r_d^2) = (2 \eta/\Gamma_n) (R_m/R_e l_c^2)$.
Using the fact $(2 \eta/\Gamma_c) (R_m/R_e l_c^2) = 1$, we
have $b = (\Gamma_c/\Gamma_n) > 1$ for all modes.

Further, since $r_d << l_c$ in general, 
in evaluating the integral in Eq.\ (\ref{hpsolo}) it will also suffice to take 
the form of $T_{LL}(r)$ in the region $0 < r < l$ to get
$(\kappa_N(0)/\kappa_N(y)) = (1 + 2 b y^2/3)^{-1}$. 
Also, we can model 
the scale dependent $\alpha_N(r)$ as follows:
At any scale $r > l_c$, in the galactic context,
the helicity structure function will scale as
\begin{equation}
2C(0) -2C(r) \sim (v_r^2 \tau_r/h)(\Omega \tau_r)  = 
(\Omega/h) r^2 ,
\label{hel}
\end{equation} 
where $v_r$ and $\tau_r$ are the eddy velocity and correlation time at any
scale $r$ which staisfy the relation $v_r \tau_r = r$. Also 
$C(r)$ is continuous at $r=l_c$, should have zero slope at the origin, and
should vanish for $r> L$.
A model helical correlation function which 
satisfies all the above requirements is given by
\begin{equation}
2C(r) = {\Omega L^2 \over h} [ 1 - {r^2 \over L^2} ] \quad {\rm for} \ 
0 < r < L \ ; C(r) = 0 \quad {\rm for} \ r > L .
\end{equation} 
So in the kinematic regime we have 
\begin{equation} 
{\alpha_N(y) \over 2 \kappa_N(0) } =  {\Omega \over 2\kappa_N(0) h} y^2 a^2 
=  {\Omega \over \Gamma_n h} y^2 .
\end{equation}
Using the above forms of  $\alpha_N(y)$, $\kappa_N(y)$ and $M_n(y)$ 
in Eq.\ (\ref{hpsolo}) we get
\begin{equation}
H_P(0,t) = C_1(b)
 {\Omega \over \Gamma_n h} M_n(0) \exp(\Gamma_n t) \label{hpfin}
\end{equation}
where $C_1$ is given by the integral
\begin{equation} 
C_1(b)= {1 \over \sqrt{6}} 
\int_0^\infty dy   
{\exp ( - \int_0^y { d y^{\prime} \over y^{\prime}}
 \left[( {9\over 4} + {y^{\prime 2}\over 1 + 2b y^{\prime 2}/3 })^{1/2}
- {3\over 2} \right] \over ( {9\over 4} + { y^2\over 1 + 2 b y^2/3 })^{1/4} } 
\times {1\over y^3}{d\over dy}
\left(y^4  {d \over  dy}(y^2 \exp(-by^2))\right) 
\label{ci}
\end{equation} 

The constant $C_1$ can be comupted numerically for any $b$.
We give below an approximate analytical estimate. 
First we note that the integrand
is suppressed exponentially for $y >> 1$. It then suffices to expand
the square root in the exponential, assuming 
$ y^2/(1 + 2by^2/3) << 9/4$. The exponential term then gives
a term $(1 + 2by^2/3)^{-1/4b}$ to the integrand of Eq. \ (\ref{ci}) .
Consider first the case $b=1$, appropriate
for the fastest growing mode.
After repeated integration by parts we then have
\begin{equation}
C_1(1) \approx {5\over 27} \int_0^\infty dw {w^3 \exp(-w^2) 
\over (1 + 10w^2/9)^{5/4} } +  {125 \over 243 } \int_0^\infty dw 
{w^5 \exp(-w^2) \over (1 + 10w^2/9)^{9/4} }
\label{ciapp}
\end{equation}
Note firstly that $ C_1(1) > 0$. Now
$y^n e^{-y^2}$ is strongly peaked about $y = (n/2)^{1/2}$
for large $n$ One can then replace the remaining slowly varying parts
in the integrands above by its value at the peak. 
The remaining integrals are gamma functions,
which can be evaluated to give $
C_1 \approx 0.05$ for for the fastest growing mode. 

For modes which extend up to 
$r \sim L$, with growth rate $\sim V/L$,  $b = R_e^{1/2} \sim 170$.
For $b >> 1$, $1/4b <<1$ and the contribution to the integrand in 
(\ref{ci}) from the exponetial term is $\approx 
(1 + 2by^2/3)^{-1/4b} \sim 1$. Also for $b>>1$, we have 
main contribution to the integral coming from regions $y^2 \sim b^{-1}$;
In this case an approximate evaluation of
the integral along lines similar to (\ref{ciapp}) gives 
$C_1(b) \approx 10^{-2} b^{-2}$, for large $b >>1$.

A number of comments are in order: Firstly, note that
in the kinematic regime, $H(0,t)$ grows exponentially.
So the assumption of small-scale stationarity
made for example by Gruzinov and Diamond (1994) to derive a 
constraint on $H$ is not valid. Also since $C_1$ is positive,
the growth of small-scale fields and hence $H(0,t)$ goes 
to decrease the effective alpha
effect. The extent of the decrease depends on how much the
small-scale field grows before it saturates.
In the kinematic regime, a mode with growth rate $\Gamma_n \sim v_l/l$,
for example, leads to a reduction in the alpha effect to a value
\begin{equation}
\alpha_{eff} = 2C(0) - 4 a H(0,t) = {\Omega L^2 \over h}
\left[ 1 -  { 4 a C_1(b_n) M_n(0) e^{\Gamma_n t} \over \Gamma_n L^2 }
\right] = 2 C(0)
\left[ 1 -  ({l \over L})^2
{ 2C_1(b_n) \over R_{ambi}(l,t)} \right] .
\label{kinal}
\end{equation}

Since $C_1 << 1$ in general and $R_{ambi}(l) >> 1$ from Eq. (\ref{rambii}), 
we see that the reduction to the $\alpha$-effect
due to ambipolar drift is negligible even as the
magnetic field energy grows to equipartition. 
A similar result obtains if we use the reduction
in alpha effect estimated in the quasilinear
approximation of Gruzinov and Diamond (1994).
In our notation their
$\alpha_{eff} = \alpha_{GD} = 2C(0) - \tau H(0,t)/(\pi \rho_n)$,
where $\tau $ is a turbulence correlation time. Using 
$H(0,t)$ determined above, one then gets
for a mode with growth rate $\Gamma_n \sim v_l/l$,
\begin{equation}
\alpha_{GD} = 2C(0)[ 1- (4C_1/3) f(l,t)(v_l \tau l)/L^2].
\label{algd}
\end{equation}
Here we have assumed that the small scale 
field energy density is a fraction $f(l,t)$ of the energy density
associated with eddies of scale $l$.
Again for $C_1 << 1$, the correction to alpha in the quasilinear
regime is modest, as long as $f(l,t) < 1$. Ofcourse, it is unclear what is 
the domain of validity of the quasilinear
approach. One should take the estimate in Eq. (\ref{algd}), 
of alpha effect reduction at best as a rough guide,
since neither we (nor Gruzinov and Diamond 1994)
are not treating the full nonlinear MHD problem. Finally note that
$H = C_1 (\Omega/\Gamma_n) M_L/h << M_L/h$, since $ (\Omega/\Gamma_n) << 1$
and $C_1 << 1$. So our neglect of the $\alpha_N$ coupling
term for the evolution of $M_L$, in section 5 is an even better
approximation than was argued there.

\section{Non-linear effects due to current helicity }

As mentioned earlier the effect 
of ambipolar drift is to add to the diffusion
coefficient an amount $ 2aM_L(0,t)$ 
and decrease $\alpha_N$ by $4aH(0,t)$, in both the mean field
and the correlation function equations. We now consider some implications
of the additional term due to the current helicity $H(0,t)$.
First let us look at the non linear evolution of $H(r,t)$ itself.
The general solution of \ (\ref{mheq})\  for the evolution of $H$, when 
$M_L$ is also settling towards a saturated state is 
beyond the scope of this work. The following general 
comments can however be made: the increase in $\kappa_N$ due to 
ambipolar diffusion will lead to a further damping of $H$. However the 
change (decrease) in $\alpha_N$ can lead in principle to 
a further increase in $H$ due to the fact that $M_L$ has a negative
curvature near the origin. 

To illustrate this more interesting 
nonlinear effect, we consider the following simpler problem. We assume 
that $M_L$ has attained a saturated state and also 
ignore the effects of turbulent diffusion for analytical
tractability. The equation for $H$ then becomes
\begin{equation} 
[{\partial \over \partial t} -2\eta_{ambi} 
{1\over r^4}{\partial \over \partial r}
\left(r^4  {\partial \over \partial r}\right)] H 
= 
{1\over r^4}{\partial \over \partial r}
\left(r^4  {\partial \over \partial r}(
\alpha(r) M_L(r) - 4aH(0,t) M_L(r))\right) \equiv  
\rho(r,t) \label{hsim}
\end{equation}
where $\eta_{ambi} = \eta + 2aM_L(0)$ as before and 
$\alpha(r) = 2C(0) - 2C(r)$.
One can formally solve this equation using the Greens 
function for the operator on the LHS of Eq.\ (\ref{hsim}).
One gets 
\begin{equation}
H(r,t) = \int_0^t dt^{\prime} \int_0^{\infty} r^{\prime 4} dr^{\prime}
G(r,t;r^{\prime},t^{\prime}) \rho(r^{\prime},t^{\prime})  +
\int_0^{\infty} r^{\prime 4} dr^{\prime}
G(r,t;r^{\prime},0) H(r^{\prime},0) \label{gsol}
\end{equation}
The Greens function $G$ can be obtained by standard 
methods (see for example Burton 1989); we get 
\begin{equation}
G(r,t;r^{\prime},t^{\prime})  = \theta(\tau) {1\over (4\eta_{ambi}\tau)^{5/2}}
\exp[-{(r^2 +r^{\prime 2}) \over 8\eta_{ambi}\tau}] 
{I_{3/2}(rr^{\prime}/(4\eta_{ambi}\tau))
\over (rr^{\prime}/(4\eta_{ambi}\tau))^{3/2}}
\label{greenn}
\end{equation}
where we have defined $\tau = t -t^{\prime}$ and 
$I_{3/2}(x) = (2/(\pi x))^{1/2} [ \cosh x - \sinh x/x] $ is the modified
Bessel function.
One can see that the Green function $G$ is similar to that of the
diffusion equation ; infact it is akin to the radial Green 
function for diffusion in 5-spatial dimensions.
We are again particularly interested in the evolution of the 
current helicity $H(0,t)$. To look at its evolution
take the limit $r\to 0$ in Eq.\ (\ref{gsol})  to get 
\begin{eqnarray}
H(0,t) &=&  {1\over 3} ({2\over \pi})^{1/2}
\int_0^t dt^{\prime} \int_0^{\infty} r^{\prime 4} dr^{\prime}
 \theta(\tau) {1\over (4\eta_{ambi}\tau)^{5/2}}
\exp[-{(r^{\prime 2}) \over 8\eta_{ambi}\tau}] 
\rho(r^{\prime},t^{\prime}) \nonumber\\   &+&
 {1\over 3} ({2\over \pi})^{1/2}
\int_0^{\infty} r^{\prime 4} dr^{\prime}
 {1\over (4\eta_{ambi} t)^{5/2}}
\exp[-{(r^{\prime 2}) \over 8\eta_{ambi} t.}] 
H(r^{\prime},0) 
\label{gsolo}
\end{eqnarray}

An initial distribution of $H(r,0)$ will be damped by diffusion.
Also since the  term proportional to $\alpha(r)$ in 
$\rho(r,t)$ is constant in time the genetration of $H(0,t)$ due to 
this term will also be damped eventually by diffusion. The only potentially
interesting term for the further growth of $H$ is the nonlinear 
term in $\rho(r,t) \propto H(0,t)$.
To see its effect, let us assume that the steady state  
longitudinal magnetic correlation function can be described by
$M_L(r) = M_L(0) \exp[-r^2/(2L_m^2)]$.
We also take an intial 
$H(r,0) = H(0,0) \exp[-r^2/(2L_H^2)] $. Using Eq.\ (\ref{gsolo}) , 
doing the $r^{\prime}$ integrals we then get 
\begin{equation}
H(0,t) = {20aM_L(0) \over L_m^2} 
\int_{0}^t dt^{\prime} 
 \theta(t-t^{\prime}) {H(0,t^{\prime}) \over (1 + 
4\eta_{ambi}\tau/L_m^2)^{7/2}}
  + {H(0,0) \over (1 + 4\eta_{ambi} t /L_H^2)^{5/2}}
\label{hot}
\end{equation}

This is an integral equation for $H(0,t)$. An approximate solution
to this equation can be obtained as follows. Consider times 
$0 < t < t_1$ such that $4\eta_{ambi} t_1 /L_m^2  < 1$.
For these times one may approximate the integrand in Eq.\ (\ref{hot}) 
by just $H(0,t^{\prime})$. Then $H(0,t)$ satisfies the simple
differential equation 
\begin{equation}
dH(0,t)/dt = 20aM_L(0) \left[ { H(0,t) \over L_m^2} - 
{H(0,0) \over L_H^2}\right]
\label{hfirs}
\end{equation}
So if $ L_m < L_H$, $H(0,t)$ can grow exponentially. We saw in section 5
and 6 that $L_m$ is generally the diffusive scale. Since during
linear evolution of $H(r,t)$, the current field correlations are
generated from $M_L(r,t)$ through the alpha effect, we expect $L_H \sim L_m$.
To estimate the maximum further growth of the current helicity,
assume that we do have $L_m < L_H$, initially. Then for $t < t_1$
the current helicity grows exponentially
with $H(0,t) = H(0,0)[(1-f_{mH}) \exp( 20aM_L(0)t/L^2) + f_{mH}] $,
where $f_{mH}= L_m/L_H < 1$. At time 
$t_1$ we then have 
$H(0,t_1) = H(0,0)[(1-f_{mH}) e^{5/2} + f_{mH}]$, where
we have used the fact $\eta_{ambi} = \eta + 2aM_L(0) \sim 2aM_L(0)$ . At  
later times, one can treat $H(0,t^{\prime})$ in the integrand of
Eq.\ (\ref{hot}) as a slowly varying function compared to 
$1/( 1 + 4\eta_{ambi}(t-t^{\prime})/L_m^2)^{7/2}$, pull it out of the 
integral and integrate the resulting equation to get 
\begin{equation}
H(0,t) = 
H(0,t_1)\left(2 - {1\over ( 1 + 4\eta_{ambi}(t-t_1)/L_m^2)^{5/2}}\right)
\label{nsol}
\end{equation} 

At large times one sees that $H(0,t) \to 2 H(0,t_1) < 24 H(0,0)$.
The upper limit obtains only when $f_{mH} << 1$.
A further growth of $H(0,t)$ in the nonlinear stage
implies a further reduction of
$\alpha_{eff}$. Incase of purely ambipolar drift this
reduction is still very small (cf. Eq. (\ref{kinal}), while 
$\alpha_{GD}$ reduction depends on how large $f(l,t)$ grows
before attaining the saturated state. For example if
as argued in Paper I, $f(l,t) << 1$ in the saturated
state then $\alpha_{GD}$ will also suffer only a modest
reduction, due to small-scale dynamo action.

Before ending this section, we mention two other interesting
non linear effects, which obtains under the action of
ambipolar drift, which both invlove the existence of
a non-zero magnetic helicity:
\begin{itemize}
\item{ (a) Relaxation through selective decay}

Suppose we start off with a random field configuration which 
has an initial non-zero magnetic helicity $P_H(0,t_i)$, and ask how it will
evolve under the action of ambipolar drift.
Due to ambipolar drift and the 
resulting ion-neutral friction, the field energy is constantly drained
into heat. This is reflected by the diffusion terms in the 
equations for $M_L$ and $H$. However, for a large conductivity, the
magnetic helicity is almost conserved, reflecting the
fact that the field is almost frozen into the ions. 
A non-zero magnetic helicity
also implies neccessarily a minimum non-zero field energy.
So the random field cannot evolve into a zero energy configuration,
but rather into a minimum energy configuration conserving
magnetic helicity. Such a selective decay of energy, conserving 
helicity is thought to lead to self organisation into
larger scale structures (Taylor 1974).
Our formalism incorporating 
ambipolar drift offers a simple route to study the dynamics
of this relaxation process, and the final relaxed state. 
A study of such relaxation
trough selective decay, using the equations for $M_L$ and
$H$ derived here is in progress and will be reported elsewhere.

\item{(b) Inverse cascade due to the non linear dynamo} 

Suppose we had no kinetic helicity. But we had
created large, random, small-scale fields, and also some
small-scale magnetic (and current) helicity. For example,
large random magnetic fields, correlated on small scales
may be generated during phase transitions in the early universe.
Further, in some of these phase transitions,
like the electroweak one, 
there are speculations that large magnetic (and current) helicities may arise
(Cornwall 1997, Joyce and Shaposhnikov 1997).
Then our model non linear equation for the mean field (\ref{meanfin}),
(and the work of Pouquet {\it et al.} 1976), indicates
that the small-scale current helicity can lead to large-scale dynamo
action. The alpha effect will be purely magnetic. This
will lead to a coupling of small scale to large scale and
"inverse cascade" (or dynamo growth) of magnetic energy
to larger scales, due to purely nonlinear
effects of the Lorentz force. If there were no other source of
magnetic energy, the energy and current helicity 
would decay montonically. However the approximate 
conservation of magnetic helicty under near ideal MHD conditions,
would still keep this dynamo active.
It would be interesting to explore this issue 
further, using the full MHD equations,

\end{itemize}

\section{ Discussion and conclusions}

We have revisited here the dynamics of fluctuating magnetic
fields in turbulent fluids, ab initio. In doing this we have
also incorporated the effects of ambipolar drift,
as would obtain in a significantly neutral gas. Ambipolar
drift introduces a magnetic field dependent addition to the velocity
field in the induction equation. The
resulting non linear equation may also be viewed, albiet with some
caution, as a toy model for the MHD problem. Assuming that
the velocity field has a turbulent component, we have derived the evolution 
equations for the mean and fluctuating magnetic field. These equations 
are used to discuss a
number of astrophysically interesting problems.
We summarise below the principle results of our work.

First, in the presence of ambipolar drift, 
the dynamo equation for the mean field, and the
equations for the magnetic correlations are both
modified. Assuming a Gaussian closure, one gets an  
extra diffusion term proportional to 
the energy density in the fluctuating fields and a reduction to
the alpha effect due to the average current helicity of the 
fluctuating fields. These equations ( Eq.\ (\ref{meanfin}) , 
Eq.\ (\ref{mleq}) and Eq.\ (\ref{mheq}) ) form a closed 
set of nonlinear partial differential
equations for the evolution of both the mean magnetic 
field and the magnetic fluctuations, incorporating the 
back reaction effects of ambipolar drift. Due to
the Gaussian closure approximation, 
the nonlinearity appears as time-dependent
co-efficients involving only the average properties of the fluctuating 
field itself.

We applied our equations in section 5 and 6
to discuss small-scale dynamo action
in galaxies, assuming a Kolmogorov type galactic
turbulence. In the kinematic phase, dynamo action exponentiates
small-scale fields provided the MRN
associated with the turbulence, exceeds a critical value 
$ R_c \approx 60$. Further, for Kolmogorov type turblulence, 
the critical MRN for excitation of a
mode extending upto $r\sim l$ is also $R_m(l) = R_c$, as
expected from the scale invariance in the inertial range.
In galaxies, it is likely that the ratio of the magnetic to fluid
reynolds number $R_m/R_e >> 1$. In this case the fastest growing
modes grow on a timescale comparable to the turn-over time of
the smallest eddies at the cut-off scale $l_c$, 
with $\Gamma \sim VR_e^{1/2}/L \sim v_c/l_c$. Modes whose longitudinal 
correlation function extends upto $r\sim l > l_c$ grow at a slower rate
$\Gamma \sim v_l/l$. Further, the field is strongly peaked about
a region $r =r_d < l_c (R_m/R_e)^{-1/2}$ about the origin for all the modes.
For the most rapidly growing mode, $w(r)$ 
changes sign accross $r=l_c$ and rapidly decays
with increasing $r/l_c$. Interms of the
Zeldovich rope-dynamo, one may picture, the small-scale field  
as being concentrated in ropy structures with thickness $\sim r_d$ and
curved on a scale upto $\sim l$ for a mode extending to $r\sim l$.

As the small-scale fields grow, ambipolar drift
adds to the diffusion
coefficient and the effective MRN, $R_{ambi}$ for
fluid motion on any scale of the turbulence decreases.
If $R_{ambi}(L)$ decreases to a value $R_c \sim 60$, 
dynamo action will stop completely. 
In a {\it sufficiently weakly ionised 
medium } ambipolar drift by itself 
can then lead to a saturation of small-scale fields
before the turbulence is drastically modified. The required
small ionisation fraction may obtain in the universe just after recombination
and in proto stellar disks, but not in galactic gas. 
The effective magnetic Reynolds number in galactic gas, 
even including ambipolar
diffusion, is much larger than $R_c$. In this case,
as the the small-scale field grows in strength, it
continues to be concentrated into thin ropy structures,
as in the kinematic regime. In a companion paper
(Subramanian, 1997 : Paper I)
we have built upon the results obtained in sections 5 and 6,
to discuss in detail how the small-scale dynamo may saturate in
the galactic context, 
while preserving large-scale dynamo
action. The crucial property of the small-scale dynamo
generated field which allows this is its spatial
intermittency. The field can build up locally
to a level which will lead to small-scale dynamo saturation,
while at the same time having a sub-equipartition
{\it average} energy density, so that the
diffusive property of the turbulence is not drastically affected.

We also considered in sections 7 and 8 the evolution of the
current-field correlations. The dynamics of 
helical correlations has
not been as extensively studied in the 
literature as that of longitudinal correlations (and the small-scale dynamo).
In the kinematic phase the average current helicity
associated with the small-scale field, $H$, decays due to diffusion,
unless forced to grow by coupling to the exponentially
growing longitudinal magnetic correlations. 
We showed in section 7, using a WKBJ approximation
to the relevant Greens function, that
the coupling with $M_L$, leads to an exponential 
growth of the current helicity $H(0,t)$
in the kinematic regime. 
So the assumption of small-scale stationarity
made for example by Gruzinov and Diamond (1994) to derive a 
constraint on $H$ is not valid.
The growth of small-scale fields and hence $H(0,t)$ goes 
to decrease the effective alpha
effect. The extent of the decrease depends on how much the
small-scale field grows before it saturates.
In general we find the reduction in the 
alpha effect due to the ambipolar drift itself,
as given by Eq. (\ref{kinal}) to be negligible
in galaxies. The alpha effect reduction due to the
growing small scale field, in the quasi linear
approximation to the full MHD employed by 
Gruzinov and Diamond (1994) is also small 
(Eq. (\ref{algd}) for $\alpha_{GD}$), 
as long as $f(l,t) < 1$.

In section 8 we considered the non linear
growth of the current helicity. 
We find that in principle $H(0,t)$ can grow further,
due to its nonlinear coupling with $M_L$. This could
lead to further reduction of $\alpha_{GD}$. Even then, 
provided the small-scale dynamo saturates at sub-equipartition levels
as argued in Paper I, the alpha effect is likely to
be preserved until the large scale field itself grows
to near equipartition level. Ofcourse, it is unclear what is 
the domain of validity of the quasilinear
approach, and so this result must be treated with
caution, until rederived in a full MHD treatment. 
 
The equations for the magnetic correlations 
incorporating ambipolar drift derived here, may offer
a simple route to study the dynamics
of relaxation through selective decay.
In the case when magnetic helicity (without 
kinetic helicity) is initially present, non linear effects
due to the Lorentz force 
can also lead to a magnetic alpha effect and
dynamo generation of large scale fields,
as envisaged in the early work of Pouquet {\it et al.} (1976).
It will be worthwhile to study these issues further,
possibly with a numerical treatment of the equations derived
here. It is also desirable to return to the dynamics
of small scale fields in the full MHD context;
Perhaps by finding simple ways of 
incorporating the dynamics of the velocity
correlations on an equal footing as that of the magnetic
correlations. We hope to 
return to these issues in the future.

%
%

\acknowledgments

At Sussex KS is supported by a PPARC
Visiting Fellowship. He thanks John Barrow for invitation to Sussex.
Leon Mestel and T. Padmanabhan are 
thanked for detailed comments on earlier drafts of 
the manuscript. Some progress on the paper was made while KS 
visited Princeton University Observatory, supported by 
NSF grant AST-9424416. He thanks Jerry Ostriker and Ed Turner 
for the invitation
to visit Princeton University Observatory, and the staff there
for a warm reception. 
Partial travel support came from IAU Commission 38.
Discussions with collegues too
numerous to mention have been very helpful, during the course of this work.
In particular KS thanks T. Padmanabhan and M. Vivekanand at Pune
for being critical but encouraging listeners.

\appendix

\section[]{ Evolution of fluctuating field correlations}

We give below more algebraic details  
involved in simplifying the first term in Eq. \ (\ref{meq} ) .
This term is given by
\begin{equation}
<\int \ ^yR_{jpq}\left(v_T^p({\bf y},t) \ ^xR_{ilm}(v_T^l({\bf x},s) 
M_{mq})\right ) ds > =
 -\epsilon_{itu}\epsilon_{ulm}\epsilon_{jrs}\epsilon_{spq}
{\partial^2 \over \partial r^r \partial r^t }\left [ T^{lp} M_{mq} \right ]
\label{appone}
\end{equation}

For examining the evolution of $M_L$ one needs to multiply the
above equation by $r^i r^j/r^2$. We can simplify the resulting equation
by using the identity
\begin{equation}
r^i r^j 
{\partial^2 A \over \partial r^r \partial r^t }
= {\partial^2 (A r^i r^j )\over \partial r^r \partial r^t }
-\delta_{jt}r^i {\partial A \over \partial r^r }
- \delta_{ir}r^j {\partial A \over \partial r^t }
-\delta_{jt} \delta_{ir} A
\end{equation}
where $A = T^{lp} M_{mq}$. Then using 
$\epsilon_{itu}\epsilon_{ulm} 
= \delta_{il}\delta_{tm} - \delta_{im}\delta_{tl}$, and the
definition of $T_{LL}, T_{NN} $ and $C$, straightforward algebra
gives the contribution of the first term to 
$(\partial M_L / \partial t)$
\begin{equation}
{\partial M_L \over \partial t}( 1 \ {\rm st} \ {\rm term}) = 
-{1\over r^4}{\partial \over \partial r}
(r^4 T_{LL}{\partial M_L \over \partial r}) + {G \over 2} M_L + 4 C H  
\end{equation}
The second term of Eq. \ (\ref{meq} ) gives an identical contribution.

To derive the evolution of $H$ due to these terms
multiply Eq. \ (\ref{appone}) by $\epsilon_{ijf}r^f$ .
Using the fact that the turbulent velocity and
small scale field have vanishing divergence, we 
have $M_{ij,j} = 0$ and $T_{ij,j}=0$. This allows one to
simplify the contribution from the first term to
\begin{equation}
{\partial H \over \partial t}( 1 \ {\rm st} \ {\rm term}) 
=-{\epsilon_{ijf} r^f \over 2 r^2} \left[ T_{ij,tr}M_{tr} +
T_{tr}M_{ij,tr} - T_{ir,t}M_{tj,r} - T_{tj,r}M_{ir,t} \right]
\label{beg}
\end{equation}
The first two terms on the RHS of Eq. (\ref{beg}) 
can be further simplified by noting that $\epsilon_{ijf}T_{ij} = 2Cr^f$ and 
$\epsilon_{ijf}M_{ij} = 2Hr^f$ . We have then 
\begin{equation}
-{ \epsilon_{ijf} r^f \over 2 r^2} \left[ T_{ij,tr}M_{tr} +
T_{tr}M_{ij,tr}\right] = -[ T_{LL}H^{\prime \prime} +T_{LL}^{\prime}
H^{\prime}+ {4T_{LL}H^{\prime} \over r} +
M_{L}C^{\prime \prime} +M_{L}^{\prime}
C^{\prime}+ {4M_{L}C^{\prime} \over r} ]
\label{ontw}
\end{equation}
Here prime denotes a derivative with respect to $r$.
To evaluate the contribution of
the last two terms on the RHS of Eq. (\ref{beg}) 
it is convenient to split up the tensors $M_{ij}$ and $T_{ij}$
into symmetric and
antisymmetric parts (under the interchange of $(ij)$ ).
We put a superscript $S$ on the symmetric part and $A$ on
the antisymmetric part. Then we can write after some algebra
\begin{eqnarray}
&&{\epsilon_{ijf} r^f \over 2 r^2} \left[T_{ir,t}M_{tj,r} + 
T_{tj,r}M_{ir,t} \right] = {\epsilon_{ijf} r^f \over  r^2} 
\left[T_{ir,t}^SM_{tj,r}^A  + T_{ir,t}^AM_{tj,r}^S \right] 
\nonumber\\ && =
-\left[ HT_{LL}^{\prime \prime} + C M_L^{\prime \prime} +
T_{LL}^{\prime}H^{\prime}+M_{L}^{\prime}C^{\prime}
+{4HT_{LL}^{\prime} \over r} + {4CM_{L}^{\prime} \over r} \right]
\label{thfo}
\end{eqnarray}
Adding the contributions from Eq. (\ref{ontw}) and (\ref{thfo})
gives 
\begin{equation}
{\partial H \over \partial t}( 1 \ {\rm st} \ {\rm term}) 
=-{1\over r^4}{\partial \over \partial r}
(r^4 {\partial \over \partial r}[T_{LL}H + CM_L]
\label{hfir}
\end{equation}
The second term of Eq. \ (\ref{meq} ) gives an identical contribution.

\section[]{ The WKBJ analysis of the kinematic small-scale dynamo}

First, in order to implement the boundary condition at $r=0$, 
under WKBJ approximation, it is better to transform to a new 
radial co-ordinate $x$, where $r=e^x$.
Also to eliminate first derivative terms in the resulting equation 
we substitute $\Phi(x) = \exp(x/2) \Theta$ and get
\begin{equation}
{d^2 \Theta \over dx^2} + p(x)\Theta = 0  
\label{tevol}
\end{equation} 
where 
\begin{equation}
p(x) = {-(\Gamma + U)e^{2x} \over \kappa_N} - {1\over 4}
\end{equation}
The WKBJ solutions to this equation are linear combinations of
\begin{equation}
\Theta = {1\over p^{1/4}} \exp (\pm i\int^x p^{1/2} dx)
\end{equation}
The solutions have to satisfy the boundary conditions 
$\Theta(x) \to 0$ for $x \to \pm \infty$. One therefore
has a standard WKBJ eigenvalue problem for the determination
of $\Gamma$.
Note that as $x\to -\infty$, $p \to -9/4$ and so the WKBJ 
solutions are in the form of growing and decaying exponentials at this end.
This also the case as
$x \to + \infty$ since $p \to -\Gamma e^{2x} < 0$ for growing mode
solutions with $\Gamma > 0$. In order to match the boundary conditions
at both ends, the solution has to be the growing exponential at
$x = -\infty$ and transit to the decaying exponential as
$x \to +\infty$. This can only obtain if the $p(x)$ goes
through zeros, by $U$ becoming negative for some range of $r$.
For $U$ considered here, in general, 
$p(x)$ goes through two zeros. Suppose these occur at $x_1$ and $x_2$
with $x_1 < x_2$. Then the 
WKBJ soultions will be oscillatory in the range $x_1 < x < x_2$.
The requirement that the oscillatory solutions match on to the 
growing exponential near $x = -\infty$ and the decaying
exponential as $x \to + \infty$, gives the standard condition
(cf. Jeffreys and Jeffreys 1966, Mestel $\&$ Subramanian 1991) 
on the the eigenvalue $\Gamma$
\begin{equation}
\int_{x_1}^{x_2} p^{1/2}(x) dx  = {(2n + 1)\pi \over 2} .
\label{wkb}
\end{equation}
One also determines the eigenfunction $\Theta(x)$, under 
the WKBJ approximation, to be 
\begin{eqnarray}
\Theta &=& {A\over (-p)^{1/4}} \exp \left[\int^x_{x_1} (-p)^{1/2} dx \right]
\qquad x < x_1 \nonumber\\
&=& {2^{1/2}A\over p^{1/4}} \sin\left[ \int^x_{x_1} (p)^{1/2} dx 
+ {\pi\over 4}\right] \qquad x_1 < x < x_2 \nonumber\\
&=& {(-1)^n A \over (-p)^{1/4}} \exp\left[ -\int^x_{x_2} (-p)^{1/2} 
dx\right] \qquad x_2 < x 
\label{eigsol}
\end{eqnarray}

\subsection{ Critical MRN for the marginal mode}

Let us first use Eq. \ ( \ref{wkb} ) to determine the critical
value of $R_m = R_c$ needed for growth of the small-scale fields.
For this we have to put $\Gamma =0$ and find $R_m$ which
satisfies (\ref{wkb} ). With $\Gamma =0$, using $U$ from
( \ref{potin} ) , defining $y = r/L$ we have in the range
$ l_c/L < y < 1$, 
\begin{equation}
p(y) = {  -(9/4) -(29 R_m/54) y^{4/3} +  
(13 R_m^2/108) y^{8/3}
\over (1 + y^{4/3}R_m/3 )^2 }
\label{pyone}
\end{equation} 
We will find that $R_m= R_c$ is large enough that
one can assume $1/R_m << y^{4/3}$ in the above expression
for $p$. In this case the zeros occur approximately at
$r_1/L \sim (174/39R_c)^{3/4} = y_0$ and at $r_2/L \sim 1$. 
(Here $r_1= \exp(x_1)$ and $r_2 = \exp(x_2)$ ). 
The integral condition then becomes
\begin{equation}
\int_{x_1}^{x_2} p^{1/2}(x) dx  = \int_{y_0}^1 [ {13\over 12 } 
- {174 \over 39 R_c y^{4/3} } ]^{1/2} {dy \over y} =  {\pi \over 2}
\label{qint}
\end{equation}
where we have looked for the critical reynolds number for
the principal mode with $n=0$. The integral in Eq. \ ( \ref{qint} ) can
be done exactly and gives the condition 
\begin{equation}
{3\over 2} ({13 \over 12})^{1/2} \left [ ln [{1 + \sqrt{1 - y_0^2} \over y_0 }]
- \sqrt{1 - y_0^2} \right ] = {\pi \over 2}
\end{equation}
The solution of this equation implies a critical value for
the excitation of the marginal mode $R_m = R_c \approx 60$.
Note that $R_c$ was estimated by taking $n=0$ in Eq. \ ( \ref{wkb} ).
One can easily get also the limiting magnetic reynolds
number needed for the exitation of higher order modes.
It should also be pointed out that in Eq. ( \ref{pyone} ) for $p(y)$ , 
$y^{4/3}R_m = (r/L)^{4/3}R_m
\equiv (r/l)^{4/3}R_m(l)$ . So the above equations determening
$R_c$ are the same if we replace $(L,R_m)$ by $(l,R_m(l))$. This
shows that the critical MRN for excitation of a
mode extendinng to $r\sim l$ is also $R_m(l) = R_c$, as
expected from the scale invariance in the inertial range.
 
The actual value of the MRN at the 
outer scale is likely to be much larger than $R_c$, in galaxies. Infact
it is likely that $R_m/R_e >> 1$. 
Let us now estimate the growth rate of the fastest 
growing mode in this case. 

\subsection{ Growth rate for the fastest growing mode}

For this we fix $R_m$ and $R_e$ and 
look for the value of $\Gamma$ which satisfies Eq. \ ( \ref{wkb} ) .
For $R_m >> R_e$,the potential $U$ 
is negative at $r=L$, decreases monotonically 
from $r=L$ to $r=l_c$ and is still negative at $r=l_c$. It only starts
increasing for $r < l_c$.
The fastest growing mode is then expected to concentrate at $r < l_c$.
Also the turning point corresponding to $x = x_1$ occurs at $r < l_c$
for all the modes. So to determine the growth 
rate of the fastest growing mode
and also examine the structure of the
modes at small radius, one must adopt the form of $T_{LL}(r)$
with $r< l_c$, given in  Eq. \ ( \ref{tll} ). For $r < l_c$, one then
has $\kappa_N = \eta[ 1 + (1/3) (r/r_d)^2 ] $ and
\begin{equation}
U(r) = {V \over 3 L} R_e^{1/2} { [ 2z^{-2} - 1 - 4z^2 ]
\over [1 +  z^2 ]}
\label{potor}
\end{equation}
Here $z = r/(\sqrt{3}r_d)$ with
\begin{equation}
r_d = { l_c \over R_m^{1/2}(l_c) }
\label{scal}
\end{equation}
setting the scale over which the potential varies. 
Using this form of the potential we then have
\begin{equation}
p(z) = {A_0 z^4 -B_0 z^2 - 9/4 \over (1 + z^2)^2}
\label{pcut}
\end{equation}
for the range $0 < r < l_c$.
Here $\bar \Gamma = \Gamma/(VR_e^{1/2}/3L)$ is a normalised growth rate, 
$A_0 = 15/4 - \bar\Gamma$ and $B_0 = \bar\Gamma - 1/2$.

For the above form of $p(z)$,
there is only one real positive zero $z_0$. It turns out
that $z_0 $ is large enough such that the $9/4$ in $p(z)$ above
can be neglected compared to the other terms giving  
$z_0 \approx (\bar\Gamma - 1/2)^{1/2}/(15/4 -\bar\Gamma)^{1/2}$.
Note that for the form of $T_{LL}(r)$ we have adopted the potential
is discontinuous at $r=l_c$. For $r= l_c -\epsilon$,
$U =U^{-} = -4 (VR_e^{1/2}/3L)$ while for $r= l_c + \epsilon$,
$U = U^{+} = -4/3 (VR_e^{1/2}/3L)$. As we mentioned earlier this
does not alter the  results qualitatively since they are 
generally dependent on integrals over $U$. It means however that for
$ 4/3 <\bar\Gamma < 4$, which we will see obtains for the
fastest growing mode, the outer zero of $p$ is at $r=l_c$ or $z=z_c =
 l_c/(\sqrt{3}r_d) = (R_m/3R_e)^{1/2}$. 
For $\bar\Gamma$ corresponding to 
the fastest growing mode, we then obtain the integral condition
\begin{equation}
\int_{z_0}^{z_c}
{[A_0 z^4 -B_0 z^2]^{1/2} \over z^2}
{dz\over z} = {\pi \over 2}
\end{equation}
In the above we have again assumed that $z_0^2 >> 1$, which
we will show to be true below.
This integral can be done exactly and gives the condition
\begin{equation}
A_0^{1/2} {\rm ln}[ 2A_0z_c^2/B_0 - 1] - 2 (A_0z_c^2 -B_0)/z_c = {\pi \over 2}.
\end{equation}
In the above condition, since $z_c^2 = (R_m/3R_e) >> 1$, one can 
get a good iterative approximate solution for $\bar\Gamma$.
In the first approximation one neglects the constant terms
compared to the $z_c^2$ terms to get
\begin{equation}
\bar\Gamma \approx 15/4 - (\pi/2ln(R_m/R_e))^2 .
\label{fastgro}
\end{equation}
For example for $R_m/R_e = 10^{15}$ one has $\bar\Gamma \sim 3.748$
or $\Gamma \sim 1.25 VR_e^{1/2}/L$. 
So as advertised the fastest growing
modes grow on a timescale comparable to the turn-over time of
the smalles, cut-off scale eddies, if $R_m >> R_e$, 
with $\Gamma \sim VR_e^{1/2}/L$. 
One can also go back and check, using this value of
$\Gamma$ that $z_0 \sim 30 >> 1$ and so 
the approximations made assuming $z_0, z_c >> 1$ are very good.

\subsection{ Spatial structure of the eigen modes}

We now briefly consider the spatial structure for various 
eigenmodes of the small-scale dynamo, 
for the case $R_m(l_c) =R_m/R_e >> 1$. 
In this case as we mentioned earlier,
the turning point corresponding to $x = x_1$ occurs at $r < l_c$
for all the modes. The eigenfunction for $r<l_c$ are then given by
Eq. ( \ref{eigsol} ) with $p$ given by the form in (\ref{pcut} ) .

Infact near the origin, one can use the 
the original equation (\ref{phievol}) 
to find the behavior of $M_L$ and $w(r)$.
One finds that for small $r$, $\Phi$ satisfies
the equation
\begin{equation}
{d^2 \Phi \over dz^2} - {2\Phi \over z^2} + \alpha \Phi = 0
\label{phizer}
\end{equation}
where the constant $\alpha = (5 -\bar\Gamma)$. 
The solution can be found by elementary methods 
$\Phi = z^2 {\rm sin}(\sqrt{\alpha} z)/(\sqrt{\alpha} z) $. So then have
\begin{equation}
M_L(r,t) = { M_L(0,t) \over [1 + z^2]^{1/2} }
{ {\rm sin}(\sqrt{\alpha} z) \over \sqrt{\alpha} z }
\label{mlorg}
\end{equation}

A similar result can also be found from the WKBJ solution,
where one gets for $ z < z_0$,
\begin{equation}
M_L = { e^{\Gamma t} F(z) \over z^{3/2} (1 + z^2)^{1/2}}
\end{equation}
with
\begin{equation}
F(z) = {A_1 \over(-p(z))^{1/4}}
 \exp\left[\int^z_{z_0} (-p(z))^{1/2} {dz \over z} \right]
\label{ozo}
\end{equation}
As $ z \to 0$, $-p(z) \to 9/4 -\alpha z^2$ and so this soln goes over to 
\begin{equation}
M_L =  { M_L(0,t) \over [1 + z^2]^{1/2} }{1 \over (1 - (4\alpha/9)z^2)}
\exp[{-\alpha z^2 \over 6}]
\end{equation}
This is in good agreement with the exact soln determined for small $z$.
For larger $z$, $M_L$ decreases monotonically with $z$. The WKBJ
treatment gives for $ z_0 < z  < z_2$ ($z_2$ is the outer turning point)
\begin{equation}
F(z) = {2^{1/2} A_1 \over(p(z))^{1/4}}
 \sin\left[\int^z_{z_0} (p(z))^{1/2} {dz \over z} + {\pi\over 4}\right]
\label{zozc}
\end{equation}
Away from $z=z_0$, the $z^4$ term in $p(z)$ dominates and 
$p \to (15/4 - \bar\Gamma)= A_0$. So for $z >> z_0$ we have
\begin{equation}
M_L \approx { e^{\Gamma t} 2^{1/2} A_1 \over A_0^{1/4} z^{5/2} }
\sin \left[A_0^{1/2} ln({z \over z_0}) + {\pi \over 4} \right]
\end{equation}
The above equations show that $M_L$ decreases rapidly with increasing
$z$. Since $z= r/r_d$, one sees that $M_L$ and hence $w(r)$ for 
all the modes are
strongly peaked about the radius $r \sim r_d = l_c/R_m^{1/2}(l_c) << l_c$,
for the case $R_m(l_c) >> 1$.

We also mentioned that $w(r)$ must become negative
at some radius. From the WKBJ solution it is apparent that
the number of zero crossings for the WKB solution will depend
on the order $n$ of the mode. Let us consider the fastest growing
mode. We saw earlier that for this mode the outer turning 
point is at $r=l_c$ or $z=z_c$. For this mode one can check that
from the WKBJ solution that $w(r) > 0$, in the region $ 0 < r < l_c$.
For $ r > l_c$ and $R_m/R_e >> 1$, 
one has from Eq. \ ( \ref{potin} ) , $U(r) \approx 
-(4 V R_e^{1/2}/9L) (r/l_c)^{-2/3} $ and 
$\kappa_N \approx (VL/3R_e)(r/l_c)^{4/3}$. Therefore 
\begin{equation}
p(r) \approx {13 \over 12} - ( {\Gamma \over (v_l /3l)}) ({r\over l})^{2/3}
\quad {\rm for} \ l_c < r < L ; \ R_m/R_e >> 1
\label{prap}
\end{equation}
For the fastest growing mode one can igmore the constant term
and get $-p(r) \approx \bar\Gamma(r/l_c)^{2/3}$. Then 
\begin{equation}
M_L(r) \propto ({l_c \over r})^{7/3} \exp\left[-3\bar\Gamma^{1/2}
((r/l_c)^{1/3} - 1)\right ]
\end{equation}
and 
\begin{equation}
w \propto {1\over y^2} \left({2\over 3}({l_c\over r})^{1/3} 
- \bar\Gamma^{1/2}\right)
\exp\left[-3\bar\Gamma^{1/2}
((r/l_c)^{1/3} - 1)\right ]
\end{equation}
For $\bar\Gamma \sim 3.75$ we can see that $w(r) < 0$ for $r>l_c$ and
its modulus decreases to zero rapidly with increasing $r/l_c$.
So $w$ changes sign accross the transition point $r=l_c$ for the
form of longitudinal correlation function $T_{LL}$ we have adopted.
For modes with a smaller growth rate $\Gamma$, we 
see from Eq. \ (\ref{prap}), that the outer turning point,
got by putting $p(r) = 0$, is at $r/l  \sim (13/(12\bar\Gamma_l))^{3/2}$,
where $\bar\Gamma_l = \Gamma/(v_l/3l)$. So modes with
growth rate $\Gamma \sim v_l/l$, extend upto $r \sim l$.

In summary, in the case $R_m/R_e >> 1$, $w(r)$ is strongly peaked about
a region $r < l_c (R_m/R_e)^{-1/2}$ about the origin for all the modes.
For the most rapidly growing mode, $w(r)$ 
changes sign accross $r=l_c$ and rapidly decays
with increasing $r/l_c$. (If one had adopted a sufficiently
smooth form for $T_{LL}(r)$ around $r=l_c$, the value where
$w$ changes sign would still be $r \sim l_c$, but could have
been better determined by the WKBJ analysis). Also
slower growing modes with $\Gamma \sim v_l/l$, extend upto $r \sim l$.
A more thorough analysis of the eigenfunctions
can be found in Kleeorin et. al. (1986), for the simple case when
the longitudinal
velocity correlation function has only a single scale. Their analysis
is also applicable to the mode near the cut-off scale of 
Kolmogorov type turbulence. 

Let us now consider the corresponding eigenfunction for the 
marginal mode. In this case, as we saw earlier, 
the inner turning point occurs at
$r = r_1 = Ly_0 \sim 0.14 L$ and the outer turning point is at
$r= r_2 \sim L$. The WKBJ solution can be used to determine the
the eigenfunction. We get for $r < r_1$,
\begin{equation}
\Theta(y) = {A_2\over (-p(y))^{1/4}} \exp (\int^y_0 (-p(y^{\prime}))^{1/2} 
{ dy^{\prime} \over y^{\prime}})
\end{equation}
where $p(y)$ is given by Eq. ( \ref{pyone} ) . For $r << L$, we can
neglect the $y^{8/3}$ term in  have Eq. ( \ref{pyone} ) compared to the
constant and $y^{4/3}$ terms. This  
implies $\Theta(y) \approx y^{3/2} \exp( -13R_my^{4/3}/54)$ and 
so for the marginal mode 
\begin{equation}
M_L(r,t) \propto { \exp( -( 13/54) R_m y^{4/3}) \over 
( 1 + (1/3)R_m y^{4/3})^{1/2} }
\label{mlmar}
\end{equation}
One sees that the eigenfunction is concentrated in a
radius $r \sim L /R_c^{3/4} = r_c$ for the marginal mode.
Infact if we  define $\bar z^2 = R_cy^{4/3}/3$, then for the marginal
mode with $\Gamma = 0$ we have from Eq. ( \ref{pyone} ),
$p = (-9/4 -25\bar z^2/18 + 13\bar z^4/12)/(1+\bar z^2)^2$.
This is very similar to the $p(z)$ defined in Eq. ( \ref{pcut} )
except for the identifications $ z \to \bar z$, $A_0 \to 13/12$,
$B_0 \to 25/18$. Going through the same analysis as
for the $p(z)$ of Eq. ( \ref{pcut} ) , gives the properties of
of $w(r)$ for the marginal mode : For the marginal mode
$w(r)$ peaks within $\bar z \sim 1$, corresponding to a
radius $r \sim L /R_c^{3/4} $, changes sign to become
negative at $r \sim L$ and dies rapidly for larger $r/L$.

\end{document}